\documentclass[pdftex,twocolumn]{aastex62}
\usepackage{color}

\newcommand{\HI}{H\,{\sc i}}
\newcommand{\CIV}{C\,{\sc iv}}
\newcommand{\SII}{S\,{\sc ii}}
\newcommand{\NaI}{Na\,{\sc i}}
\newcommand{\SiII}{Si\,{\sc ii}}

\def\apj{ApJ}
\def\apjs{ApJS}
\def\mnras{MNRAS}
\def\nat{Nat}
\def\apjl{ApJ}
\def\aap{A\&A}
\def\aj{AJ}

\def\apss{Ap\&SS}

\def\asec{\ifmmode ^{\prime\prime}\else$^{\prime\prime}$\fi}

\def\grad{$^\circ$}
\def\it{\sl}
\def\degs{\ifmmode ^{\circ}\else$^{\circ}$\fi}
\def\amin{\ifmmode ^{\prime}\else$^{\prime}$\fi}
\def\asec{\ifmmode ^{\prime\prime}\else$^{\prime\prime}$\fi}
\def\fss{\hbox{$.\!\!^{\rm s}$}}        
\def\fdg{\hbox{$.\!\!^\circ$}}          
\def\farcs{\hbox{$.\!\!^{\prime\prime}$}}  
\def\h{$^{\rm h}$}
\def\m{$^{\rm m}$}
\def\s{$^{\rm s}$}
\def\j1731{J1731$-$4744}
\def\b1727{B1727$-$47}
\def\degs{\ifmmode ^{\circ}\else$^{\circ}$\fi}
\def\amin{\ifmmode ^{\prime}\else$^{\prime}$\fi}

\received{January 1, 2018}
\revised{January 7, 2018}
\accepted{\today}
\submitjournal{ApJ}

\shorttitle{PSR~B1727$-$47 and RCW~114} \shortauthors{Shternin et al.}

\begin{document}


\title{
Tracking the footprints of the radio pulsar B1727$-$47: proper motion, host supernova remnant, and the glitches}


\correspondingauthor{P.~Shternin}
\email{pshternin@gmail.com}

\author[0000-0002-5810-668X]{P.~Shternin}
\affiliation{Ioffe Institute, Politekhnicheskaya
 26, St.~Petersburg, 194021, Russia}

\author[0000-0002-8139-8414]{A.~Kirichenko}
\affiliation{Instituto de Astronom\'ia, Universidad Nacional Aut\'onoma de M\'exico, Apdo. Postal 877, Ensenada, Baja California,
M\'exico, 22800}
\affiliation{Ioffe Institute, Politekhnicheskaya
 26, St.~Petersburg, 194021, Russia}

\author[0000-0002-8521-9233]{D.~Zyuzin}
\affiliation{Ioffe Institute, Politekhnicheskaya
 26, St.~Petersburg, 194021, Russia}

 \author[0000-0002-7208-1706]{M.~Yu}
 \affiliation{National Astronomical Observatories, Chinese Academy of Sciences, Beijing, China}

\author{A.~Danilenko}
\affiliation{Ioffe Institute, Politekhnicheskaya
 26, St.~Petersburg, 194021, Russia}

\author[0000-0002-4931-4612]{M.~Voronkov}
\affiliation{CSIRO Astronomy \& Space Science, PO Box 76, Epping, NSW 1710, Australia}

\author{Yu.~Shibanov}
\affiliation{Ioffe Institute, Politekhnicheskaya
 26, St.~Petersburg, 194021, Russia}

\begin{abstract}
The bright radio pulsar B1727$-$47 with a characteristic age of $80$~kyr
 is among the first  pulsars discovered 50 yr ago.
Using
its regular  timing observations  and  interferometric positions at three  epochs,
we  measured, for the first time, the pulsar proper motion of $151 \pm 19$~mas~yr$^{-1}$.
At the dispersion measure distance of $\ga 2.7$~kpc, this would  suggest a record
 transverse velocity of the pulsar
$\ga1900$~km~s$^{-1}$. However,
a backward extrapolation   of the pulsar track to its birth epoch points remarkably  close to the center
of the evolved
 nearby supernova remnant RCW~114, which suggests genuine association of the two objects. In
this case, the pulsar is substantially  closer ($\sim 0.6$~kpc)    and
younger ($\sim 50$~kyr), and its velocity  ($\sim400$~km~s$^{-1}$)
is compatible with the observed pulsar velocity distribution. We also identified two new glitches of the pulsar.
We discuss implications of our results on  the pulsar and  remnant properties.
\end{abstract}

\keywords{ISM: individual objects (RCW 114) -- ISM: supernova remnants -- pulsars: general -- pulsars: individual (PSR~\j1731)}

\section{Introduction}
 \label{S:intro}
Associations of pulsars (PSRs) with  their supernova remnants (SNRs) provide crucial information
on  the properties of these astrophysical objects
born at the same supernova explosions.
In particular, this allows one to get the most reliable constraints on objects'
ages, local environments,
and on supernova progenitor parameters.
 Justified
associations are useful for the study of  the whole PSR--SNR population
which is important for unveiling the routes of the final stages
of the stellar evolution \citep[e.g.,][]{2003heger}.

Relatively short lifetimes of the extended remnants ($\la100$~kyr) limit the  identification
of each pulsar with its SNR. So far only about fifteen radio PSRs have been firmly identified with SNRs \citep[e.g.,][]{yao2017}.
The number of candidate associations is permanently growing  due to ongoing progress
in the observational instrumentation. However,
a significant amount of them
could be due to a chance spatial coincidence \citep{Gaensler1995MNRAS}.
Therefore, each possible PSR--SNR connection
requires  several independent justifications.
Among them, the highest priority is granted to  PSR proper
motion (p.m.) measurements \citep[e.g.,][]{Kaspi1996ASPC}. Generally,
a pulsar must
be kicked
from
its parent supernova explosion center.
If the p.m.~track of the pulsar misses the remnant
center by a significant distance, this can be considered as a strong argument against the association \citep[e.g.,][]{Brisken2006ApJ},
although sometimes the off-centric asymmetric solutions are possible
\citep[e.g.,][]{Gvaramadze2002}.

\begin{deluxetable*}{*{6}{l}}[t]
\tablecaption{Published B1727 coordinates\label{tab:published_pos}}
\tablehead{
\colhead{Epoch}  & \colhead{R.A.,  B1950} & \colhead{Decl., B1950} & \colhead{R.A., J2000} & \colhead{Decl., J2000}
 & \colhead{Telescope,}  \\
 \colhead{MJD, yr month } & \colhead{} & \colhead{} & \colhead{} & \colhead{} & \colhead{Reference}
 }
\startdata
41638,  1972 Nov &  17\h27\m56\s(1) & $-$47\degs42\amin22\asec(2) & 17\h31\m42\fss6(1.0) & $-$47\degs44\amin33\farcs7(2.0) & Molonglo, 1 \\
 43494, 1977 Dec  & 17h27\m55\fss38(2) & $-$47\degs42\amin21\farcs4(5) & 17\h31\m41\fss99(2) & $-$47\degs44\amin33\farcs0(5) & Tidbinbilla, 2\\
47780.5,  1989 Sep & 17\h27\m55\fss42(6) & $-$47\degs42\amin23\asec(1) & 17\h31\m42\fss06(6) & $-$47\degs44\amin34\farcs7(1.0) & Mt Pleasant, 3 \\
 50059,  1995 Dec&  &  & 17\h31\m42\fss103(5) & $-$47\degs44\amin34\farcs56(14) & Parkes,  4 \\
54548,  2008 Mar&  &  & 17\h31\m42\fss17(7) & $-$47\degs44\amin37\asec(2) & Parkes, 5\\
\enddata
\tablecomments{
References: 1--\citet{Vaughan1974MNRAS}; 2--\citet{Manchester1983MNRAS}; 3-\citet{DAlessandro1993MNRAS}; 4--\citet{Wang2000MNRAS}; 5--\citet{Yu2013MNRAS}.
The numbers in brackets are uncertainties related to the last significant digits quoted.}
\end{deluxetable*}

One of the targets interesting  for establishing a new PSR--SNR connection by the p.m.~measurements
is PSR B1727$-$47 (J1731$-$4744; hereafter B1727). The pulsar  was discovered at the Molonglo observatory
\citep{Large1968Natur} soon after the discovery of first pulsars. Its  Galactic coordinates are $l=342\fdg57$
and $b=-7\fdg67$.
According to the ATNF pulsar
catalogue \citep{manchester2005}\footnote{\url{http://www.atnf.csiro.au/research/pulsar/psrcat} (v1.58)}, B1727  is the fourth brightest pulsar
in the radio among the relatively young ($\tau_c < 100$~kyr)  pulsars known\footnote{This is true for the 0.4, 1.4, and 2.0 GHz frequency bands.}.
It has the period $P \approx 0.83$~s, the characteristic age  $\tau_c \approx 80$ kyr, and
the spin-down energy loss $\dot{E}\approx1.1\times 10^{34}$~erg~s$^{-1}$, typical for
pulsars of such an age.
The spindown-estimated dipole magnetic field is $B\approx1.2\times 10^{13}$~G.
The dispersion measure   ${\rm DM}\approx123$~pc~cm$^{-3}$ places B1727 at the distance $D=2.7$~kpc  according
to the Galaxy electron density model
NE2001 by \citet{cordes2002astro.ph} or at 5.5~kpc that follows from the YMW16 model \citep{yao2017}. Recently, the $\gamma$-ray discovery of the pulsar with \textit{Fermi}/LAT
 was reported revealing B1727 as the slowest rotator among the  known $\gamma$-ray pulsars \citep{Smith2019}.

The pulsar is projected on the limb of  the shell-like H$\alpha$ nebula RCW~114 with an  angular diameter of $\approx$ 4\grad,
also known  as G343.0$-$6.0 in SNR catalogs \citep{Green2014, Greenrep2014}. This suggests that the association between the objects is worth investigation. Large angular separation between the RCW~114 center and the B1727 position implies that a significant proper motion can be expected in case of the association.

Neither p.m.~estimates nor counterparts in other spectral domains for B1727 have been reported so far.  Only one,  unsuccessful,
attempt to estimate the pulsar velocity by the analysis of its scintillation pattern evolution in the radio
was undertaken \citep{Johnston1998MNRAS}. Nevertheless, as B1727 has a long observational history, it is interesting to search for indications
of its p.m. in published data.

\section {The p.m.~signature from
published data}
\label{pub-data}
\begin{figure}[tbh]
  \begin{center}
   \includegraphics[width=\columnwidth,clip]{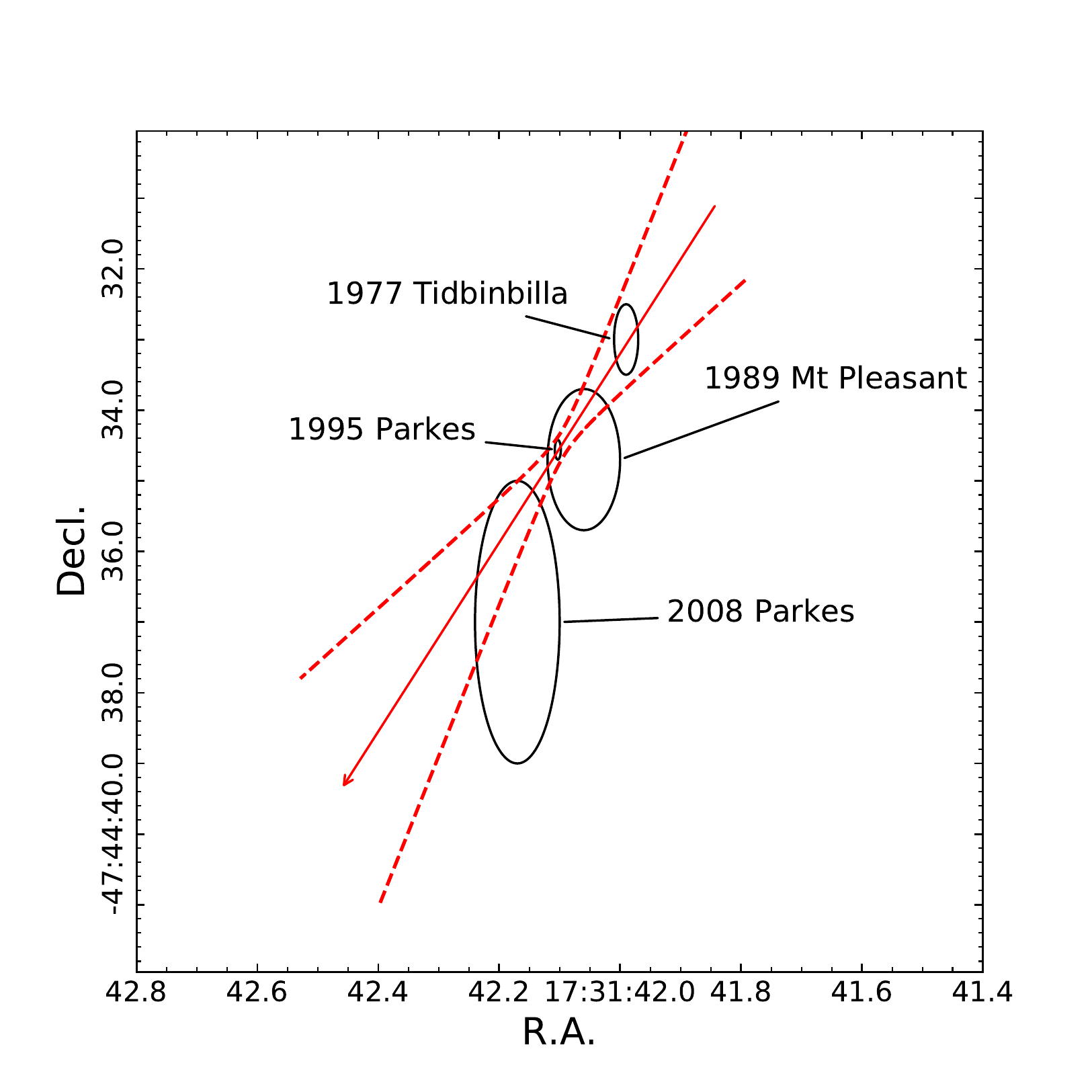}
  \end{center}
  \caption{Published positions of B1727 obtained with various telescopes at different epochs as marked in the plot.
  Here and in other plots ellipse
  sizes correspond to the 1$\sigma$ coordinate uncertainties  presented in Table~\ref{tab:published_pos} (39\% 2D probability coverage). The Molonglo position (the first line in Table~\ref{tab:published_pos}) has too large uncertainties and is not shown.
  The arrow and dashed lines
  show the p.m.~vector and its uncertainties; the vector length shows  the pulsar shift
  at a time base of  100 yr.}
 \label{fig:publ.pos}
\end{figure}
\begin{deluxetable*}{*{7}{l}}[t]
\tablecaption{Parameters of the two new glitches of B1727 found in this
work\label{tab:glithces}}
\tablehead{
\colhead{Epoch} & \colhead{$\Delta \nu_{\rm g}/\nu$} & \colhead{$\Delta \dot{\nu_{\rm g}}/\dot{\nu}$} &\colhead{$Q$} &\colhead{ $\tau_{\rm d}$} & \colhead{Number} &\colhead{ Data span }\\
\colhead{MJD,  yr month} & \colhead{$10^{-9}$} &\colhead{$10^{-3}$}&&\colhead{d}& \colhead{of ToAs} &\colhead{MJD}
}
\startdata
\hline 55735.18(14), 2011  Jun & 53.6(1.2) & 3.4(0.6)& 0.125(0.014)&141(25) & 37 & 55272--56214\\
56239.86(77),  2012 Nov  & 10.7(1.7) & 1.7(1.9)& 0.14(0.1) &70(96)&26&55897--56512 \\
\enddata
  \tablecomments{\quad$\nu=1/P$ is the pulse frequency, $\dot{\nu}$ is its first time derivative, $\Delta\nu_{\rm g}$ and $\Delta \dot{\nu_{\rm g}}$ are the changes of the frequency and
  its first derivative during the glitch,
  $\tau_{\rm d}$ is the exponential recovery time after the glitch, $Q$ is the ratio of  the transient
  frequency increment to $\Delta\nu_g$ describing the fractional glitch recovery.
 }
\end{deluxetable*}
There exist five published positions of B1727 under the B1950 and/or J2000 coordinate systems measured with various telescopes
at different  epochs between 1972 and 2008 (Table~\ref{tab:published_pos}).

As anticipated, the data    show  a noticeable positive increase
in the R.A.~and a negative trend in the decl.
This is clearly seen in Figure~\ref{fig:publ.pos}, where the
position from the first row of Table~\ref{tab:published_pos} for the epoch of 1972 is omitted  because
of
large uncertainties.
The linear
fits to $\alpha$ (R.A.) and $\delta$ (decl.) coordinate
changes
over the  epoch
yield a formally significant ``published-position''
(PP) p.m.: $\mu_\alpha^{\rm PP} = 63\pm11$~mas~yr$^{-1}$, $\mu_\delta^{\rm PP}=-83\pm 27$~mas~yr$^{-1}$, and
$\mu^{\rm PP}=104\pm 22$~mas~yr$^{-1}$ with  $\chi^2=1.6$ for 6 degrees of freedom (d.o.f.).\footnote {Hereafter  the p.m.~components are given in the Euclidean space, $\mu_\alpha=\dot{\alpha} \cos \delta$ and $\mu_\delta=\dot{\delta}$.}    The p.m.~vector is
shown by the arrow in Figure~\ref{fig:publ.pos} with uncertainties (dashed lines).
The backward extrapolation of the p.m.~using the characteristic
age of 80~kyr roughly points to the RCW~114 center. This p.m. results in  the following  transverse velocity of the pulsar
$v_{\perp}=(1300\pm 200)\times D_{2.7}$~km~s$^{-1}$, where $D_{2.7}$ is the distance in 2.7~kpc units. At the YMW16 DM distance of 5.5 kpc this is the record velocity, and the value is still quite large for the lower NE2001 DM distance of 2.7 kpc.
To
have a 2D speed in the typical for pulsars range 50$-$500~km~s$^{-1}$ \citep[e.g.,][]{hobbs2005MNRAS},
B1727  should be substantially  closer, namely at $\sim$0.5~kpc.
This is
consistent with the RCW 114 distance of $\la1.5$~kpc \citep{Kim2010ApJ} thus favoring the association with the SNR.

This
possibility, however, should be considered with
a grain of salt
as the  differences in the reported  coordinates
can be just a result of   unknown systematics between different telescopes$/$instruments.  To confirm it, detailed analyses of  homogeneous data sets are required.

The latter is possible as
B1727 has been a permanent target  of the Parkes 64\,m telescope
southern pulsar survey.
Thanks to this,
the timing analysis of the data obtained from 1990 to 2010  revealed
four glitches  in B1727 which occurred with intervals of about 2-5 yr
\citep[][and references therein]{Yu2013MNRAS}.
Unfortunately, in addition to the frequent glitches, the pulsar shows a prominent
timing noise hindering  the p.m.~measurement using a standard  timing analysis.
However, there are
series of archival interferometric observations obtained in 2004--2005 and 2011
with the Australia Telescope Compact Array (ATCA)
which
can
be
used to verify
the reliability of the timing p.m.~derivations.

In the following sections we present the
measurements of the
p.m.~of
B1727 using  the Bayesian method \citep{2009feroz,Lentati2014MNRAS} on the available archival data
collected
with the Parkes telescope until 2014. We also use the ATCA archival data and our own observations obtained in 2016.
Preliminary results of the analysis
were briefly reported by \citet{2017peter}.
Here we also present two new pulsar glitches occurred between 2010 and 2014. The rest of the paper is organized as follows.
The Parkes timing data and analysis are described in Section~\ref{Parkes-timing}. In Section~\ref{S:atca}, we present the ATCA
data and p.m.~measurements. The p.m.~results based on the published, timing, and interferometric data are concatenated in Section~\ref{S:comb} and the association with RCW~114 is
considered in Section~\ref{S:RCW114_assoc}. The results
are discussed in Section~\ref{S:Discussions} and
summarized in Section~\ref{S:Conclusions}.

\section {Parkes timing data and analysis}\label{S:timing}
\label{Parkes-timing}
The largest
data set  for the timing analysis of B1727  obtained with a single telescope is provided by the Parkes telescope archive\footnote{\url{http://data.csiro.au}}.
We selected the data obtained from  1993 February to  2014 March (MJD range
49043--56740). The detailed information on the observations
and filterbank systems is given by \citet{Yu2013MNRAS} who used
the data obtained during the period between  1990 October and 2010 November (MJD range 48184--55507) partially overlapping with our adopted MJD range.
We applied the \texttt{PSRCHIVE} tool \citep{Hotan2004}  to the archival data to obtain the pulse times of arrival (TOAs).
As a result,  222 TOAs were obtained spanning 21~yr in total.

\begin{deluxetable*}{*{6}{l}}[t]
\tablecaption{Derived  B1727 inter-glitch timing coordinates\label{tab:inter-glith-coor}}
\tablehead{
\colhead{Epoch}  & \colhead{R.A., J2000} & \colhead{Decl., J2000}& \colhead{Interval range} & \colhead{Number}  \\
 \colhead{MJD,  yr month}& \colhead{} &  \colhead{} & \colhead{MJD}  &\colhead{of TOAs}
}
\startdata
49203,   1993 Aug &  17\h31\m42\fss13(26) & $-$47\degs44\amin37\farcs9(6.7)  & 49043.81$-$49363.21 & 20\\
 50059, 1995 Dec & 17\h31\m42\fss09(2) & $-$47\degs44\amin34\farcs55(52) & 49415.05$-$50703.24 & 47 \\
51590, 2000 Feb & 17\h31\m42\fss112(37) & $-$47\degs44\amin36\farcs98(84)&  50722.19$-$52458.51 & 33\\
53018, 2004 Jan & 17\h31\m42\fss169(21)& $-$47\degs44\amin36\farcs78(49)& 52484.38$-$53553.61 & 28 \\
54659, 2008 Jul & 17\h31\m42\fss187(17) & $-$47\degs44\amin37\farcs57(39)& 53589.40$-$55730.53 & 63\\
55986, 2012 Feb &17\h31\m42\fss225(92) & $-$47\degs44\amin38\farcs0(1.8) & 55759.45$-$56214.16
&22\\
56497, 2013 Jul &17\h31\m42\fss179(31) & $-$47\degs44\amin38\farcs65(67) & 56255.06$-$56740.76 & 18 \\
\enddata
\end{deluxetable*}
\begin{figure}[tbh]
  \begin{center}
   \includegraphics[scale=0.3,
   bb=66 27 570 751,
   angle=-90,clip]{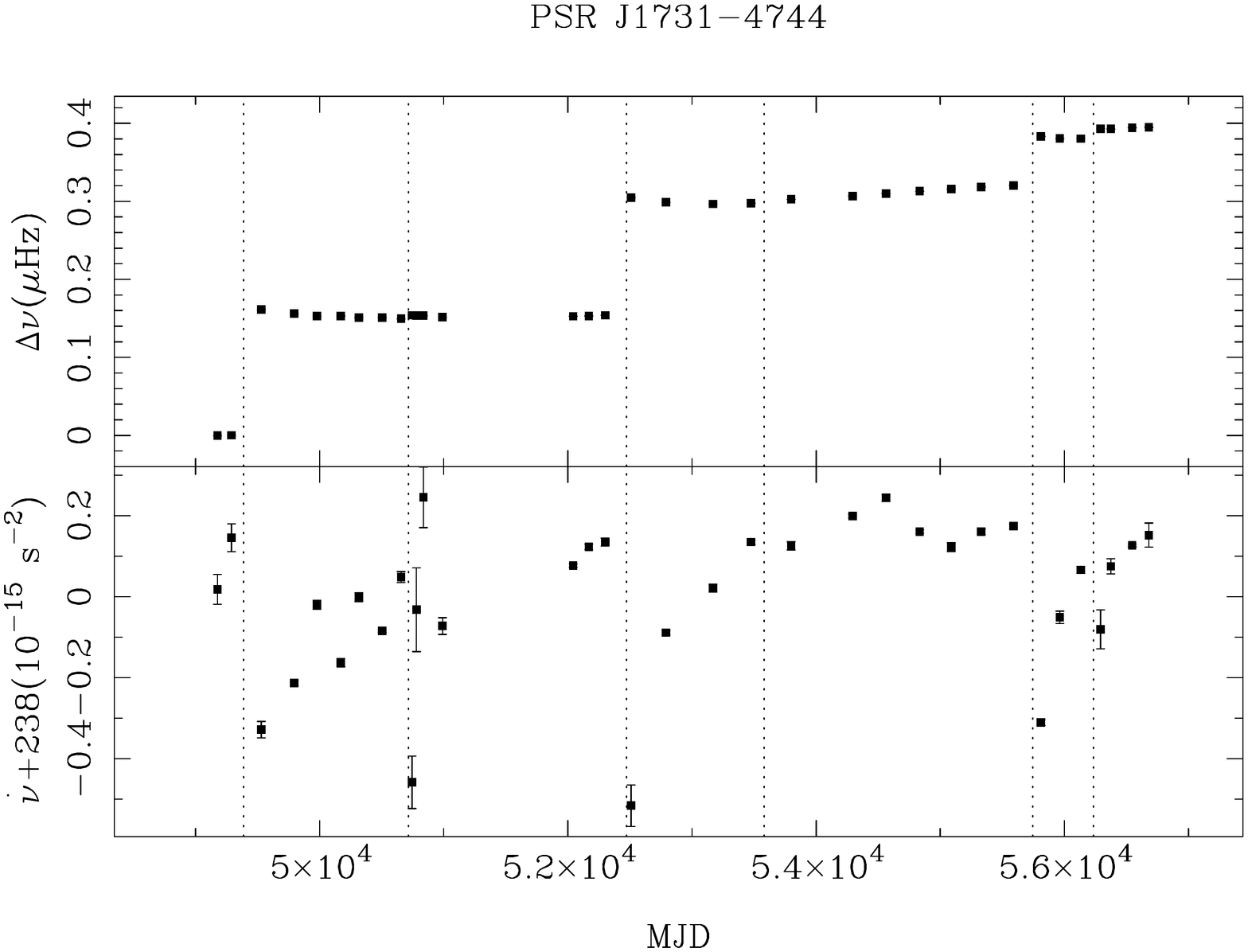}
  \end{center}
  \caption{Glitches in B1727.
  The pulse-frequency residuals $\Delta\nu$ ({\sl top}), obtained by
subtracting the (extrapolated) pulse frequency derived from the $\nu$ and $\dot\nu$ values of the first pre-glitch solution, and
 the variations of the pulse-frequency first time derivative $\dot\nu$ ({\sl bottom}).
 The glitch epochs are indicated by vertical dashed lines. The last two glitches are found in this work, while the previous
 four are from \citet{Yu2013MNRAS}.
 }\label{fig:glitches}
\end{figure}

As the considered data span includes more recent observations than in  \citet{Yu2013MNRAS}, we
first searched for new glitches in B1727.
To do that,
we used the \texttt{TEMPO2}  timing package
\citep{Hobbs2006MNRASTempo2, Edwards2006MNRASTempo2} and followed the method described by \citet{Yu2013MNRAS}.
As a result,
we found two new glitches.
Their parameters,  shown
in Table~\ref{tab:glithces}, are intermediate among those of the previous four
glitches described by \citet{Yu2013MNRAS}. Variations of the pulse-frequency residuals $\Delta\nu$ and
the pulse-frequency first time derivative $\dot\nu$ over the whole data span are presented in Figure~\ref{fig:glitches} where the positions
of all six glitches are marked by vertical lines.

With the glitch parameters in hands we turned to the inference of
astrometric parameters.  It is complicated by a
strong timing noise of B1727. Several methods were
developed to get rid of the timing noise. Recent approaches
include a generalized least squares fitting developed by
\citet{Coles2011MNRAS} and the Bayesian approach realized
in the \texttt{TEMPONEST} utility \citep{Lentati2014MNRAS}.
\citet{Li2016MNRAS} recently showed that the two approaches give consistent results, especially for
  the p.m.~measurements.
 Moreover, they
 argued that
 the novel timing analyzing techniques  give  results generally consistent with
 the interferometric measurements.  In this work, we
 employ the \texttt{TEMPONEST}  utility using two different approaches for the p.m.~derivation
 to avoid possible bias.

 In the first approach, we  selected
 inter-glitch intervals  to fit for the pulsar positions at the respective epochs.
  The resulting coordinates, epochs, interval ranges, and numbers of TOAs in the intervals are presented in Table~\ref{tab:inter-glith-coor}. Two positions (MJD 50059 and MJD 54659) are for the same interglitch intervals as the already published positions (the last two rows in Table~\ref{tab:published_pos}) and are compatible with them within the errors. Notice exclusively  small errors for the MJD 50059
 position published by \citet{Wang2000MNRAS}, which is presented in the 4-th row in Table~\ref{tab:published_pos}
 and shown in Figure~\ref{fig:radec} by the open  diamond near the relative epoch of $-$3000.
The authors applied a
``polynomial whitening''
method
suggested by  \citet{Kaspi1994ApJ}.
As has been discussed by
\citet{Coles2011MNRAS}, in this approach the uncertainties, as a rule, are strongly underestimated. This is consistent with our findings for the same data (the second row in Table~\ref{tab:inter-glith-coor}), where we get almost the same central value, but larger errorbars. At the same time, our uncertainties for the fifth interglitch interval are smaller than given in the last row in Table~\ref{tab:published_pos} because \citet{Yu2013MNRAS} had less data for that interval than is available for us now.

\begin{figure}[ht]
  \begin{center}
   \includegraphics[width=0.95\columnwidth,clip]{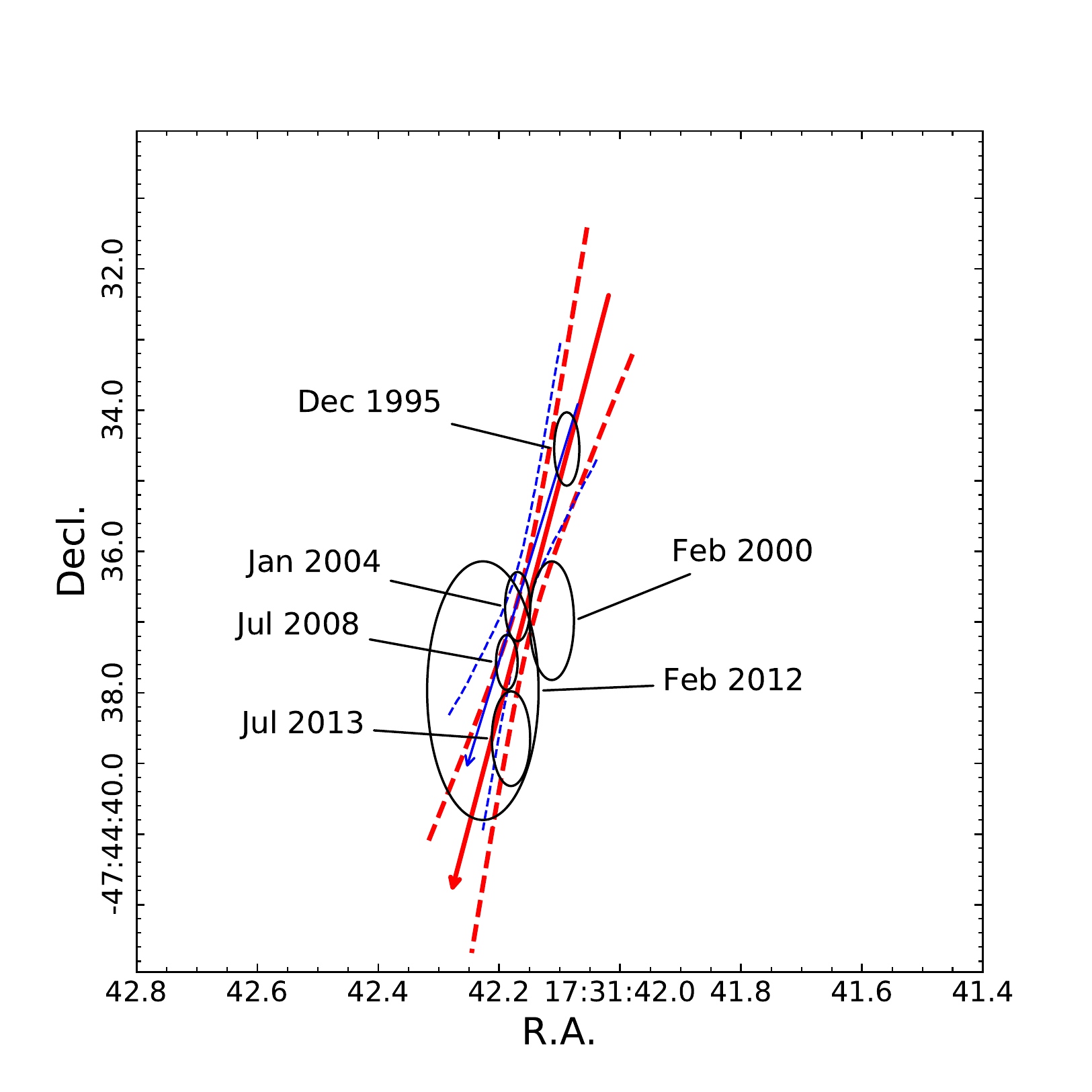}
  \end{center}
  \caption{The same as in Figure~\ref{fig:publ.pos} but for the Parkes timing positions of B1727 for six inter-glitch epochs
  listed in  Table~\ref{tab:inter-glith-coor}. The 1993 Aug position that has too large uncertainties is not shown. Thick and thin lines correspond to the p.m. measurements under the IG and GTS timing solutions, respectively, see text for details. The p.m. vector lengths for both solutions correspond here to the pulsar shift at a time base of 40 yr.
 }\label{fig:inter-glitch-pos}
\end{figure}

Based on the positions in Table~\ref{tab:inter-glith-coor}, we then derived the `inter-glitch' (IG) p.m.~using the linear  fits as in Section~\ref{pub-data}.
This resulted in $\mu_\alpha^{\rm IG}=67\pm18$~mas~yr$^{-1}$ and
 $\mu_\delta^{\rm IG}=-214\pm40$~mas~yr$^{-1}$ with $\chi^2=5.2$ for 10 d.o.f.  The respective positions and the p.m.~vector (thick arrow) are shown in Figure~\ref{fig:inter-glitch-pos}
 where the p.m.~uncertainties (thick dashed lines) are propagated  from the average position at the mean epoch of June 2000.
 As seen, this result is in line with the ``guess'' provided by the published positions (cf. Figure~\ref{fig:publ.pos}),
 although the p.m.~value $\mu^{\rm IG}=224\pm 38$~mas~yr$^{-1}$ is about twice as large as compared with
 $\mu^{\rm PP}=104\pm 22$~mas~yr$^{-1}$.
 The positional angle  of the p.m.~ vector defined from North to East
 is PA$^\mathrm{IG} = 161\degs\pm 7\degs$. It is slightly higher
 than the respective angle PA$^\mathrm{PP}=141\degs\pm11\degs$ derived from the published data.

\begin{figure}[t]
\begin{center}
   \includegraphics[scale=0.26,angle=-90,clip]{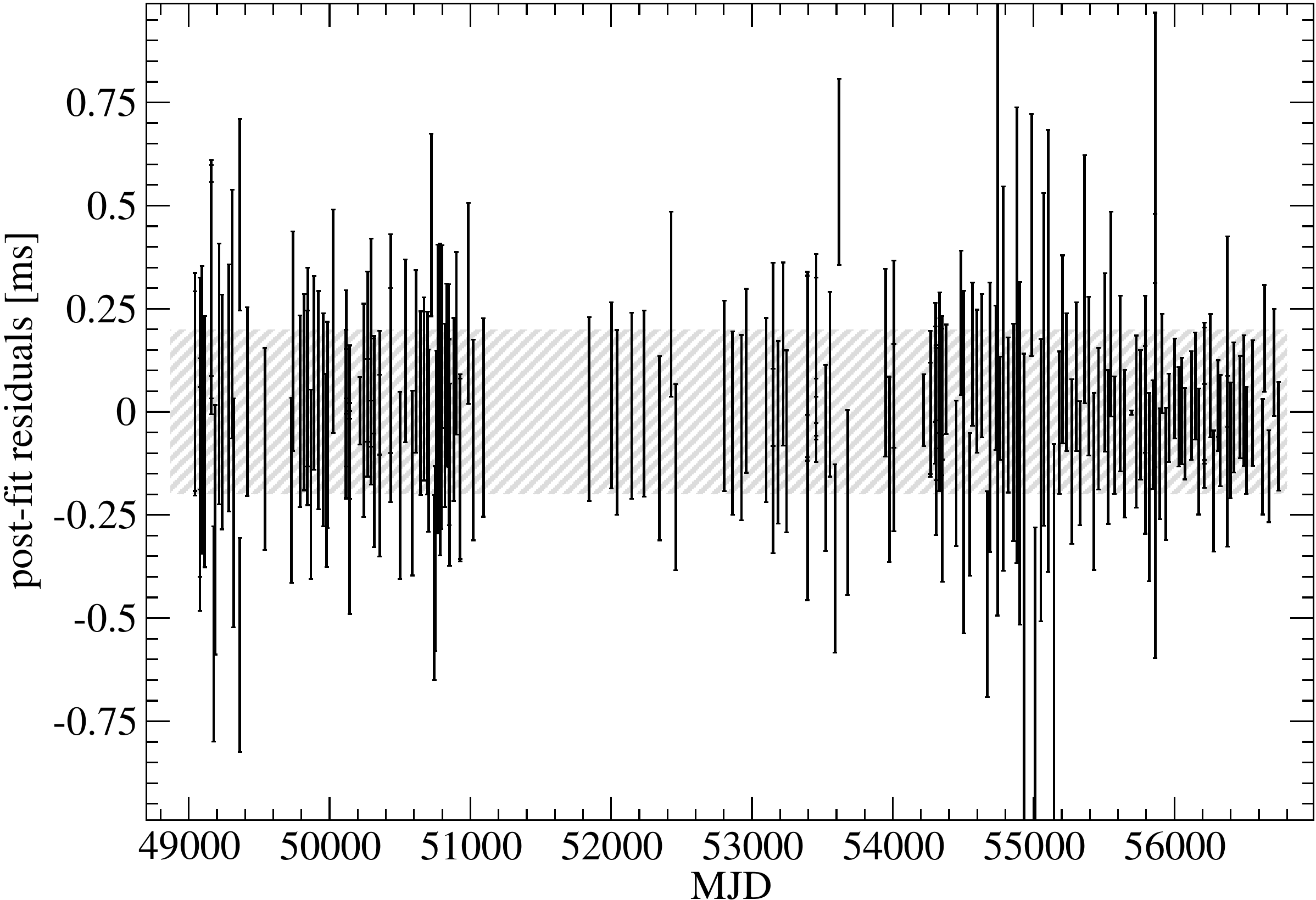}
  \end{center}
  \caption{Timing post-fit residuals of the whole data span of B1727 after taking out the timing noise with the \texttt{TEMPONEST}.
  The hatched region represents  the weighted post-fit rms residual of $\approx0.2$~ms.
 }\label{fig:t-resid}
\end{figure}

In the second approach we used the whole data span.
 The glitch parameters were, however,
 fixed in the \texttt{TEMPONEST} input as they have been derived
 at the initial analysis step.  As a result, we obtained the ``global timing-solution''  (GTS) with the
 pulsar reference position R.A.=17\h31\m42\fss14(1), decl.=$-$47\degs44\amin36\farcs71(23) at a mid-span  epoch of MJD 52892,  and  the p.m.
 $\mu_\alpha^{\rm GTS}=47\pm14$~mas~yr$^{-1}$, $\mu_\delta^{\rm GTS}=-133\pm
 37$~mas~yr$^{-1}$ at 68\% confidence.
 The weighted post-fit rms residual was $\approx202$~$\mu$s.
  Time variations of post-fit residuals  shown in Figure~\ref{fig:t-resid} demonstrate the fit quality.

The GTS pulsar track is shown with the thin arrow in Figure~\ref{fig:inter-glitch-pos} with uncertainties (thin dashed lines).  For $\mu_\alpha$, both approaches  provide compatible results within uncertainties. However, the p.m. in decl. direction is  lower in the GTS solution than that in the IG solution at 90\% confidence, resulting  in somewhat lower overall $\mu$ value for the GTS solution.
This demonstrates complications in accurate measurements of the astrometric parameters for noisy and glitching pulsars.

\section{ATCA interferometric observations and data  analysis}\label{S:atca}
The first ATCA observations of B1727 were performed during a
number of short sessions between 2004 December and 2005
March\footnote{project C1323}. We selected the data set obtained
on 2005 March 26, with the $\approx$2h15m hour angle coverage.
Although the coverage is short, this is the only session that
allows for pulsar position measurements with a reasonable
accuracy. The remaining shorter-coverage sets resulted in
significantly larger position uncertainties and were found to be
useless for our goals. The array was in the 6A configuration which
contained baselines from 337 to 5939~m. The 128-MHz band was used
with the central frequency of 1.384 GHz. In order to distinguish
the pulsed emission from B1727 and background sources and boost
signal to noise ratio for the pulsar data, the observations were
performed exploiting the correlator pulsar binning capability.

The next ATCA observations of the pulsar were obtained on 2011 November 19$-$20
as a part of The Compact Array Pulsar Emission Survey (CAPES)\footnote{project C2566}.
The observations were performed in the 16-cm band
available with the Compact Array Broadband Backend (CABB) \citep{wilson}
using the pulsar binning mode, with an hour angle coverage of about 11h.
The band had a central frequency of 2.1 GHz and covered the spectral range of 1.1$-$3.1 GHz.
The array configuration was 1.5D with baselines ranging from 107 to 4439 m.

In addition, we performed unscheduled observations\footnote{project CX367}
of the pulsar on 2016 September 15 using Director's time, that allowed for the $\approx$3h20m hour angle coverage.
The data set was obtained with the same CABB frequency setup as in the 2011 observations, however the available array
configuration H168 provided shorter baselines, from 61 to 4469 m, and no pulsar binning was possible.

\begin{deluxetable*}{*{10}{l}}[t]
\tablecaption{Interferometric coordinates of B1727 measured with the ATCA.\label{tab:t-atca-pos}}
\tablehead{
\colhead{Date}  &  \colhead{R.A., J2000} & \colhead{Decl., J2000} & \colhead{$r1$}  & \colhead{$r2$} & \colhead{PA}  &\colhead{f$_\mathrm{peak}$}& $S/N$ &\colhead{$r1_\mathrm{beam}$}  & \colhead{$r2_\mathrm{beam}$}
  \\
   \colhead{}& \colhead{} & \colhead{}  & \colhead{arcsec}  & \colhead{arcsec} & \colhead{degrees}&
 \colhead{mJy}    &   \colhead{}&\colhead{arcsec}  & \colhead{arcsec}
}
\startdata
2005 Mar 26 & 17\h31\m42\fss123(15) & $-$47\degs44\amin36\farcs200(81)  & 0.167 & 0.021 & $-$61.8
&21.7(2)& 118 &32.7&4.1 \\
2011 Nov 19$-$20 & 17\h31\m42\fss212(5) & $-$47\degs44\amin37\farcs022(67)  & 0.067
& 0.046 & 6.6
&15.0(3)&45&5.0&3.5\\
2016 Sep 15 & 17\h31\m42\fss263(9) & $-$47\degs44\amin38\farcs276(160) & 0.179 & 0.032 & 27.1 &14.5(3)&43&13.1&2.3\\
\enddata
  \tablecomments{$r1$, $r2$, and PA are the error ellipse radii  and its position angle (North to East), respectively. $f_\mathrm{peak}$ is the pulsar peak value, $S/N$ is the pulsar signal to noise ratio, and $r1_\mathrm{beam}$ and $r2_\mathrm{beam}$ are the synthesized beam semi-axes. The beam PA is equal to the PA measured for the pulsar.
 }
\end{deluxetable*}

Three data sets  described above
were processed using the
\texttt{MIRIAD} (Multichannel Image Reconstruction, Image Analysis and Display) package \citep{miriad} following standard methods and the
images were analyzed using \texttt{Karma} visualization tools \citep{gooch}.
The absolute flux density scale and bandpass calibration for all epochs were obtained from observations of PKS 1934$-$638,
whereas the gain and phase instabilities were accounted using the 1657$-$56, 1714$-$397, and 1740$-$517 calibrators for the 2005, 2011, and 2016 data sets, which lie $\approx$ 9\fdg75,  8\fdg02, and 4\fdg50 away from the target, respectively.
In the case of the 2011 data, the primary
calibration procedure revealed an unusual behavior of the \textit{YY} polarization component that was visible across
the bandwidth and
affected all baselines. The cause of the problem remained unclear, but it could probably result from an unrecognized correlator issue.
To avoid possible calibration problems, we thus used only the \textit{XX} polarization for the analysis of this data set.
\begin{figure}[t!]
  \begin{center}
   \includegraphics[width=0.95\columnwidth,clip]{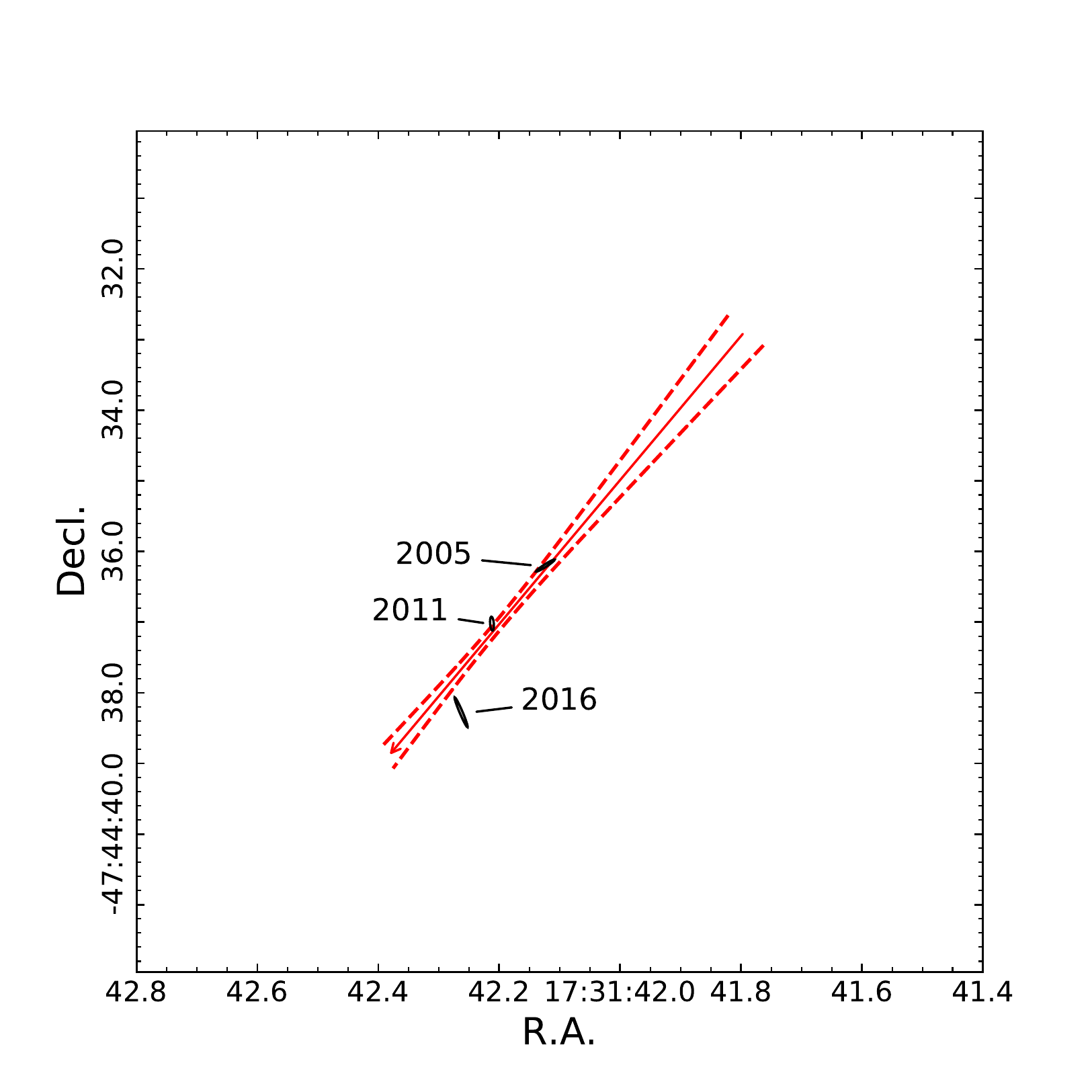}
  \end{center}
  \caption{The ATCA interferometric positions of B1727  for the three epochs listed in  Table~\ref{tab:t-atca-pos}. The p.m. track with uncertainties is shown for a 40~yr time base.
}\label{fig:f-atca-pos}
\end{figure}

The  data sets were imaged and deconvolved using standard routines.
For the 2011 and 2016 data sets
we used the \texttt{MIRIAD} \texttt{mfclean} deconvolution
tool to account for spectral variations across the  band. The  variations were neglected  in the 2005 data obtained in a much narrower bandwidth of 128 MHz and the \texttt{MIRIAD} \texttt{clean} tool was used.
The pulsar
positions were then measured on the respective full band on-pulse
images of 2005 and 2011, and  on the  full band
image of 2016
using the task \texttt{imfit}.

\begin{deluxetable*}{l*{5}{c}|*{4}c}[t]
\tablecaption{The Cartesian components of the p.m.~vector in R.A.~($\mu_\alpha$) and
decl.~($\mu_\delta$)  directions and the total
p.m.~value ($\mu$) and the position angle (PA). Different columns represent the p.m.~results obtained by different
methods. \label{tab:proper}
}
\tablehead{
\colhead{Component}  & \colhead{PP}& \colhead{IG} & \colhead{GTS} & \colhead{I} &  \colhead{I$_\mathrm{rel}$} &\colhead{F}  & \colhead{F1} & \colhead{F$_\mathrm{rel}$}&\colhead{F1$_\mathrm{rel}$}
}
\startdata
   $\mu_\alpha$, mas~yr$^{-1}$        & $63\pm11$            & $67\pm18 $  & $47\pm14$   & $150\pm10$ & $81\pm19$  & $84\pm 4$ & $87\pm 5$&$64\pm 7$&$66\pm7$\\
   $\mu_\delta$, mas~yr$^{-1}$        & $-83\pm 27$      &$-214\pm40$  &  $-133\pm  37$  & $-152\pm8$ &$-132\pm23$& $-139\pm 7$ & $-142\pm 7$&$-133\pm14$&$-147\pm14$\\
   $\mu$, mas~yr$^{-1}$ & $104\pm 22$      &$224\pm38$  &  $142\pm  34$ & $213\pm 13$ &$155\pm29$& $162\pm 8$&$166\pm 9$&$148\pm 16$&$161\pm 17$\\
   PA, deg & $141\pm11$ & $162\pm 6$ & $159 \pm 9$ & $135\pm 4$ &$148\pm10$& $149 \pm 2 $& $148 \pm 3 $&$154\pm5$&$156\pm4$\\
\enddata
  \tablecomments{Column names and methods: PP --  using the published timing positions; IG -- using the positions
  provided  by the \texttt{TEMPONEST} timing solutions for the
  inter-glitch intervals; GTS -- using the global timing solution provided by
  the \texttt{TEMPONEST} with the predefined glitch parameters; I -- using the interferometry positions;
  I$_\mathbf{rel}$ -- interferometry result in the `relative astrometry' approach;
  F (F1) -- final solution concatenating  the PP, I, and GTS (IG) results.  Notice, that the PP points used in concatenation do not include Parkes data already taken into account in GTS (IG) results to avoid double counting.
 Accordingly, F$_\mathrm{rel}$ (F1$_\mathrm{rel}$) combines the PP, I$_\mathrm{rel}$, and GTS (IG) results, see text for details. The errors correspond to  the 68\% confidence level.
}
 \end{deluxetable*}

The pulsar was firmly detected in all images with signal to noise ratios $S/N$ and peak fluxes $f_\mathrm{peak}$ given in Table~\ref{tab:t-atca-pos}. Notice the difference in $f_\mathrm{peak}$ for 1.384~GHz (2005 epoch) and 2.1~GHz (2011 and 2016 epochs) bands.
The resulting pulsar coordinates with respective
uncertainties are given in Table~\ref{tab:t-atca-pos} and the position error ellipses are shown in Figure~\ref{fig:f-atca-pos}.
According to the resulting positions, the pulsar demonstrates a significant regular shift between the 2005 and 2016 epochs which is
compatible with shifts obtained from the timing measurements (cf. Figure~\ref{fig:inter-glitch-pos}).
The  resulted ``interferometry'' (I) p.m. values are
$\mu_\alpha^{\rm I}=150\pm10$~mas~yr$^{-1}$,
 $\mu_\delta^{\rm I}=-152\pm8$~mas~yr$^{-1}$, and $\mu^{\rm I}=213\pm 13$~mas~yr$^{-1}$.
 The $\delta$-component, which dominates the overall pulsar p.m., is consistent within
 uncertainties with the results of the timing analysis.
 In contrast, the
 $\alpha$-component
 is significantly larger than those found from the timing analysis, irrespective of which solution, IG or GTS, is adopted.

In Table~\ref{tab:t-atca-pos} we quote purely statistical uncertainties which are roughly consistent with the expected ones calculated as a half of the synthesized beam size (given in the last columns in Table~\ref{tab:t-atca-pos}) divided by $S/N$. However, there can be systematic uncertainties exceeding
the statistical ones. Partially, this can be guessed from Figure~\ref{fig:f-atca-pos} where the position error ellipses seem not to follow the straight line and from a relatively large value of the $\chi^2=10$ for 2 d.o.f. for the proper motion fit. One source of the systematic errors is the absolute position uncertainties of the phase calibrators. Since three different calibrators were used, these errors are not canceled in the proper motion studies (as would have been if the same calibrator for all the epochs had been used). The calibrator position errors are negligible as compared to the pulsar position  statistical errors in the 2005 data ($\approx$0.4~mas for 1657$-$56 ) and in the 2016 data ($\approx$8$\times$4~mas for 1740$-$517) while they appear
to be noticeable for the 2011 data  ($\approx$25$\times$52~mas for 1714$-$397)\footnote{ftp://ftp.ga.gov.au/geodesy-outgoing/vlbi/Projects/The Project/astrocat.txt}. Adding the 1714$-$397
uncertainties in quadrature to the statistical ones leads only to a slight improvement of the fit ($\chi^2=8.7$), while the p.m. vector remains the same within the errors.
Another source of systematic uncertainties can be related to the calibration  and/or deconvolution algorithm errors.
The most unreliable in this respect is
the H168 array (2016 epoch) which has unbalanced baselines and is not well-suited for the astrometry tasks. The source of calibration errors can come from the antenna position uncertainties and incomplete compensation of the tropospheric/ionospheric propagation effects which are
known to increase with the separation between the calibrator and the target \citep[e.g.,][]{Chatterjee2004ApJ}.
These errors are difficult to estimate from the data as we have not found any cataloged
radio source with precisely measured coordinates in the ATCA field of view.

However, these systematic errors can be estimated by comparing the positions of the uncatalogued sources between the epochs. By doing so we found considerable shifts between the positions of most of the point sources in the field. Therefore we followed the approach of \citet{Kirichenko2015MNRAS} and performed the relative astrometry of the ATCA images using as the reference the positions of eight point sources around the pulsar firmly detected in all the images.
The details of the procedure are given in Appendix~\ref{S:atca_rel}. In short, we allowed for shifts and rotations between the images as well as for an overall scaling. The astrometric solution revealed $\approx0\farcs6$ shift in R.A. between
the 2005 and 2011 images and $\approx 0\farcs6$ shift in decl. between the 2011 and 2016 images. Moreover, the successful solution was obtained only when the 2011 image was scaled by a factor of $\approx 0.998$. The most likely
cause of the latter factor is related to the additional frequency averaging performed by CABB in the pulsar binning mode. After applying this transformation, the positions of the reference sources agreed well between the epochs (see Appendix~\ref{S:atca_rel}).
In this way we lose the information on the absolute positions of the pulsar (and the reference sources) but can estimate its proper motion. As a result, we found $\mu_\alpha^{\mathrm{I}_\mathrm{rel}}=81\pm 19$~mas~yr$^{-1}$ and $\mu_\delta^{\mathrm{I}_\mathrm{rel}}=-132\pm 23$~mas~yr$^{-1}$. While the decl. component is consistent within errors with the estimate based on the absolute (more precisely, relative to the calibrators) interferometric positions, the R.A. component decreased significantly. Notably, this `relative astrometry' p.m. estimate is consistent within 1-2$\sigma$ with the various timing-based estimates discussed above.

\section{
Concatenation of the published, timing, and interferometric p.m.~results
}\label{S:comb}
\begin{figure*}[th]
 \begin{center}
 \includegraphics[width=0.9\textwidth,clip]{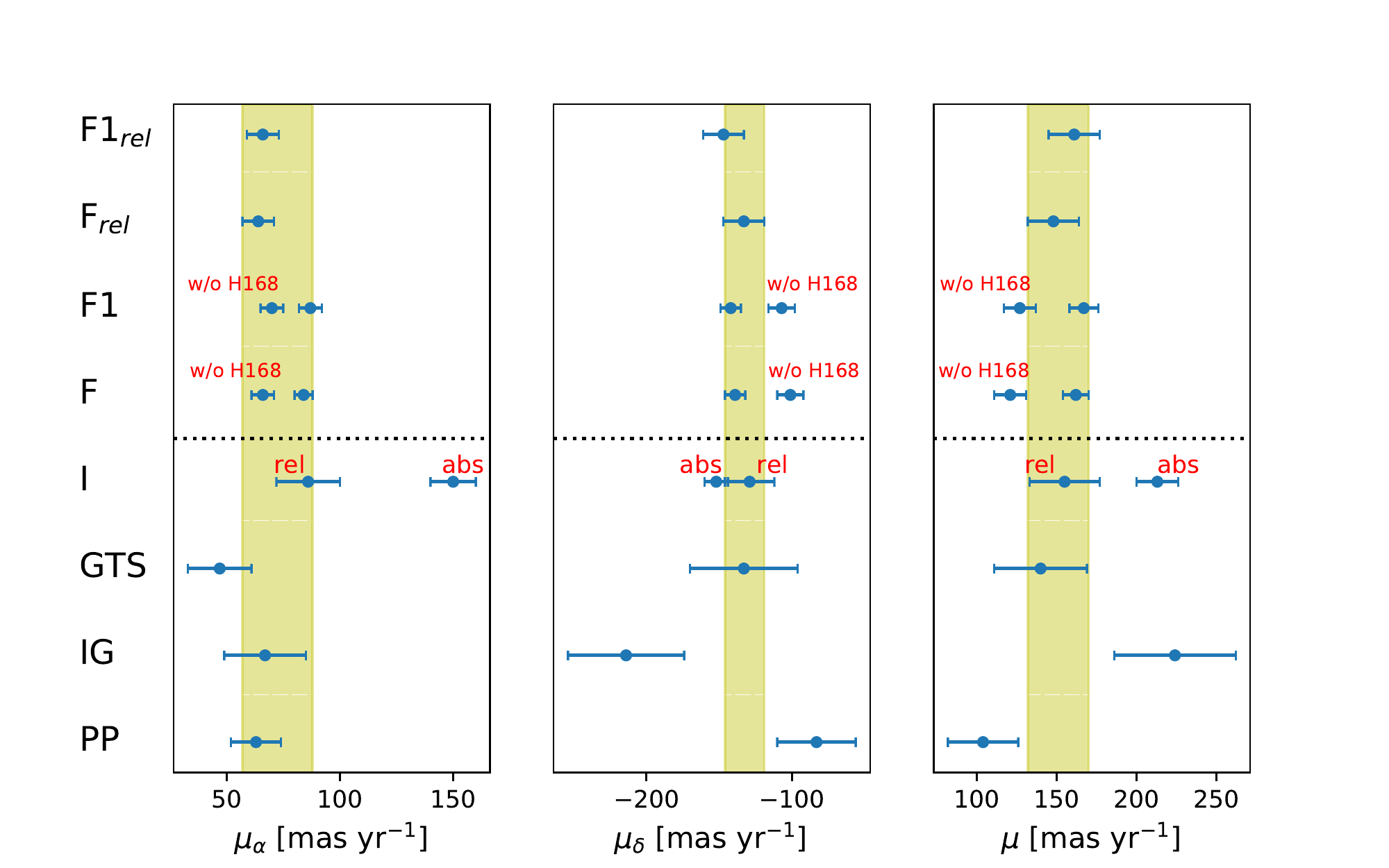}
 \end{center}
    \caption{ Summary of the proper motion measurements using various methods described in the paper. Dotted line separates the independent measurements (below) and the results from different approaches to their combination (above). Vertical strips show our final conservative p.m. estimate, see text for details.
   }
   \label{fig:Allpos}
\end{figure*}
In Table~\ref{tab:proper} and Figure~\ref{fig:Allpos}, we collect all p.m.~measurements based on  the published positions (Section~\ref{pub-data}),
 the Parkes timing positions (Section~\ref{Parkes-timing}),  and the ATCA interferometry  positions (Section~\ref{S:atca}). Despite  differences in some of the p.m. components, all approaches result in substantial proper motion roughly in the same direction. To get the most robust pulsar astrometry it is useful to consider all the data together.

\begin{figure*}[t]
 \begin{center}
 \begin{minipage}{0.45\textwidth}
 \includegraphics[width=\textwidth,clip ]{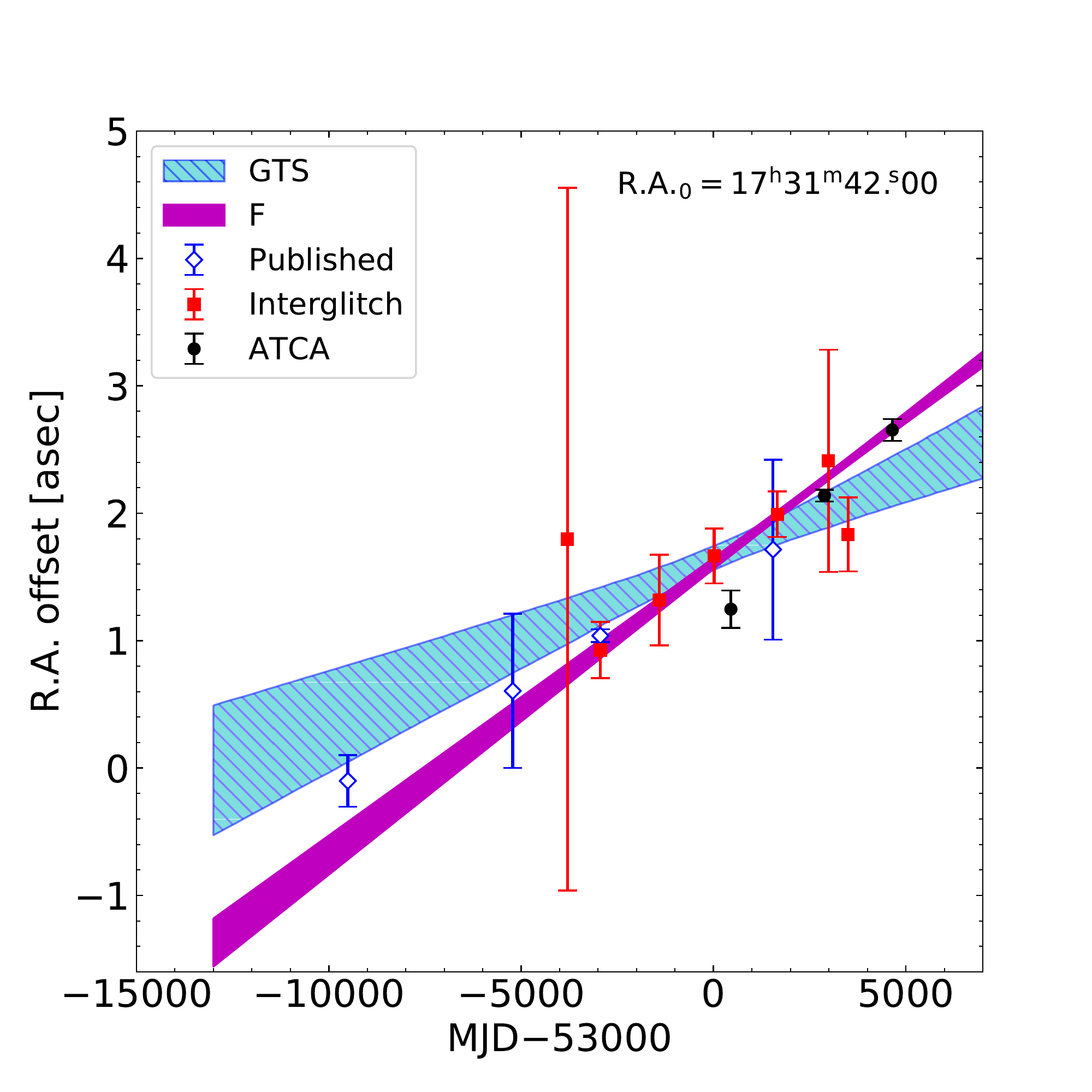}
 \end{minipage}
 \hspace{0.05\textwidth}
 \begin{minipage}{0.45\textwidth}
 \includegraphics[width=\textwidth,clip]{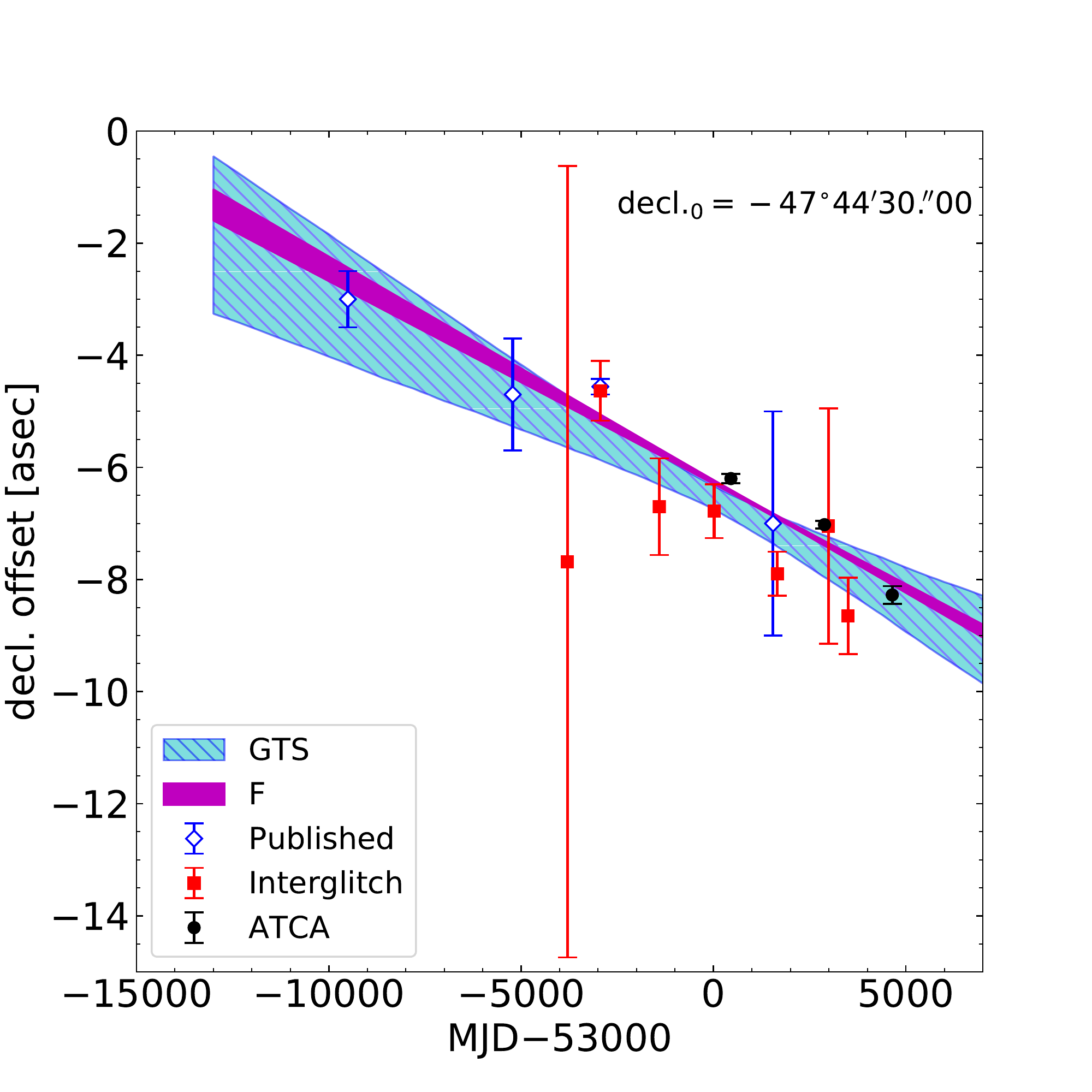}
 \end{minipage}
 \end{center}
    \caption{ B1727 R.A.~({\sl left}) and decl.~({\sl right}) variation with the epoch based on the published (open diamonds),
    and our timing inter-glitch (filled squares) and ATCA interferometric (filled circles) positions measured  at different epochs.
    The coordinates are presented in offsets relative to a reference point with RA$_0$ and decl.$_0$ shown in the plots.
    The epoch axis is referenced to  MJD 53000. Hatched and filled strips  show 68\%  credible regions
    for the pulsar p.m.~track based on our global timing solution
    and  on our
    combined solution F, respectively.
   }
   \label{fig:radec}
\end{figure*}

In Figure~\ref{fig:radec},
we plot  the  published pulsar coordinates  (Table~\ref{tab:published_pos}), inter-glitch coordinates (Table~\ref{tab:inter-glith-coor}), and interferometry coordinates (Table~\ref{tab:t-atca-pos})  by the open diamonds, squares, and circles, respectively.
The earliest position, corresponding to
the first row in  Table~\ref{tab:published_pos}, has
too large uncertainties
and is not shown.
The position epochs are  referenced to the MJD 53000, and the coordinates are shown in relative offsets from
an auxiliary reference point with R.A.$_0$ and decl.$_0$ as indicated in the plots.
Hatched strip in Figure~\ref{fig:radec} shows the 68\%  credible region for the pulsar track
based on our global timing solution obtained with \texttt{TEMPONEST} in  Section~\ref{Parkes-timing}.
As seen, all data points, including the published and interferometric ones,  are consistent with each other for the same epochs and with the GTS
region at $1-2\sigma$ level.

However, there is a clear tension between the interferometric p.m. in the R.A. direction as measured relative to the calibrators, $\mu_\alpha^\mathrm{I}$, and other measurements, which is best seen in Figure~\ref{fig:Allpos}. In this Figure, the p.m. components based on the `absolute' ineterferometric positions are
marked by `abs'. Indeed, fitting the linear pulsar track using the   interferometric   positions, the timing inter-glitch positions, and the two published positions before MJD 49043 (excluding again the Molonglo point) together, we obtain the ``F1'' solution presented in Table~\ref{tab:proper} and Figure~\ref{fig:Allpos}. However, since the interferometric fit is not good, the ``F1'' fit is not good either, with $\chi^2=76$ for 20 d.o.f. Taking out the H168 position (epoch 2016) we obtain a much better fit with $\chi^2=28$ for 18 d.o.f. ($p$-value of 6.2\%). In this case, which is marked with `w/o H168' labels in Figure~\ref{fig:Allpos} and not shown in Table~\ref{tab:proper} for brevity, $\mu_\alpha=70\pm 5$~mas~yr$^{-1}$ and $\mu_\delta=-107\pm9$~mas~yr$^{-1}$.

It is not possible to combine our favored GTS timing solution with other positions in the similar fashion, since the former results from the Bayesian analysis of the whole TOAs dataset based on the Markov Chain Monte Carlo simulations which return the sample from the posterior distributions of the pulsar position at the reference epoch and the proper motion parameters. These posteriors need to be updated with the information on the interferometric and published positions. To do this, we approximated the joined posterior distributions of the four astrometric parameters  marginalized over other model parameters with the multivariate normal distribution which works well for our TEMPONEST result. This approximated posterior was then used in combination with likelihoods for other positions assuming the pulsar linear motion to obtain the updated inference on the parameters. This resulted in the `F' proper motion solution also shown in Table~\ref{tab:proper} and Figure~\ref{fig:Allpos}. It gives  $\mu_\alpha^\mathrm{F}=84\pm 4$~mas~yr$^{-1}$ and $\mu_\delta^\mathrm{F}=-139\pm7$~mas~yr$^{-1}$, consistent with the `F1' solution based on the inter-glitch timing positions. Although it is not straightforward to extract the goodness-of-fit for this solution, we expect that the situation is similar to the `F1' case, i.e. the tension between the I positions and the GTS solution results in a bad combined fit. This is illustrated with the filled strip in Figure~\ref{fig:radec} which shows the pulsar track under the `F' solution and its 68\% uncertainty range. Its R.A. part (left panel) deviates either from the GTC and the interferometric position.
Performing the same fit without H168 position gives $\mu_\alpha=66\pm5$~mas~yr$^{-1}$ and $\mu_\delta=-142\pm 7$ as marked with `w/o H168' in the `F' row in Figure~\ref{fig:Allpos}. This is consistent with the respective `F1 w/o H168' solution.

The interferometric results I$_\mathrm{rel}$ based on the direct registration of the images for three ATCA epochs can be considered as more robust and less affected by the systematic calibration errors than the `I' results. Since the information about the absolute position is lost during the registration procedure, it is impossible to combine the results of timing and interferometric analysis based on the pulsar positions as  above. Instead, we combined the timing solution, IG or GTS, with the two published positions and then took the weighted means of the resulting p.m. component values with the respective I$_\mathrm{rel}$ components.
In this way we get the final `F$_\mathrm{rel}$' and `F1$_\mathrm{rel}$' solutions in Table~\ref{tab:proper} and Figure~\ref{fig:Allpos}, based on the GTS or IG timing results, respectively.
These two results are consistent with each other and with  timing and I$_\mathrm{rel}$ values. Based on the discussion in Sections~\ref{S:timing}--\ref{S:atca}, we suggest the F$_\mathrm{rel}$ as the most reliable proper motion solution, that accounts in the best way for the timing noise in the Parkes data and the systematics in the ATCA data.

To be more conservative, however, we will further consider the extended region which contains both the F and the F$_\mathrm{rel}$ solutions as our final p.m. estimate as indicated with the filled vertical strips in  Figure~\ref{fig:Allpos}. This range is also in agreement with the results obtained by exclusion of the H168 data point and with the results obtained by doubling the ATCA position errors (not shown for brevity).
Therefore, our conservative estimate is $\mu_\alpha=73\pm 15$~mas~yr$^{-1}$ and $\mu_\delta=-132\pm 14$~mas~yr$^{-1}$ resulting in the total p.m. $\mu=151\pm19$~mas~yr$^{-1}$ in the direction specified by the position angle $\mathrm{PA}=151\degs\pm 8\degs$. This yields  the pulsar transverse velocity  $v_{\perp}=(1930\pm 240)\times D_{2.7}$~km~s$^{-1}$.

\begin{figure*}[!ht]
 \begin{center}
     \begin{minipage}{0.45\textwidth}
     {\includegraphics[width=\textwidth,trim={3.3cm 6.5cm 2cm 6.5cm},clip]{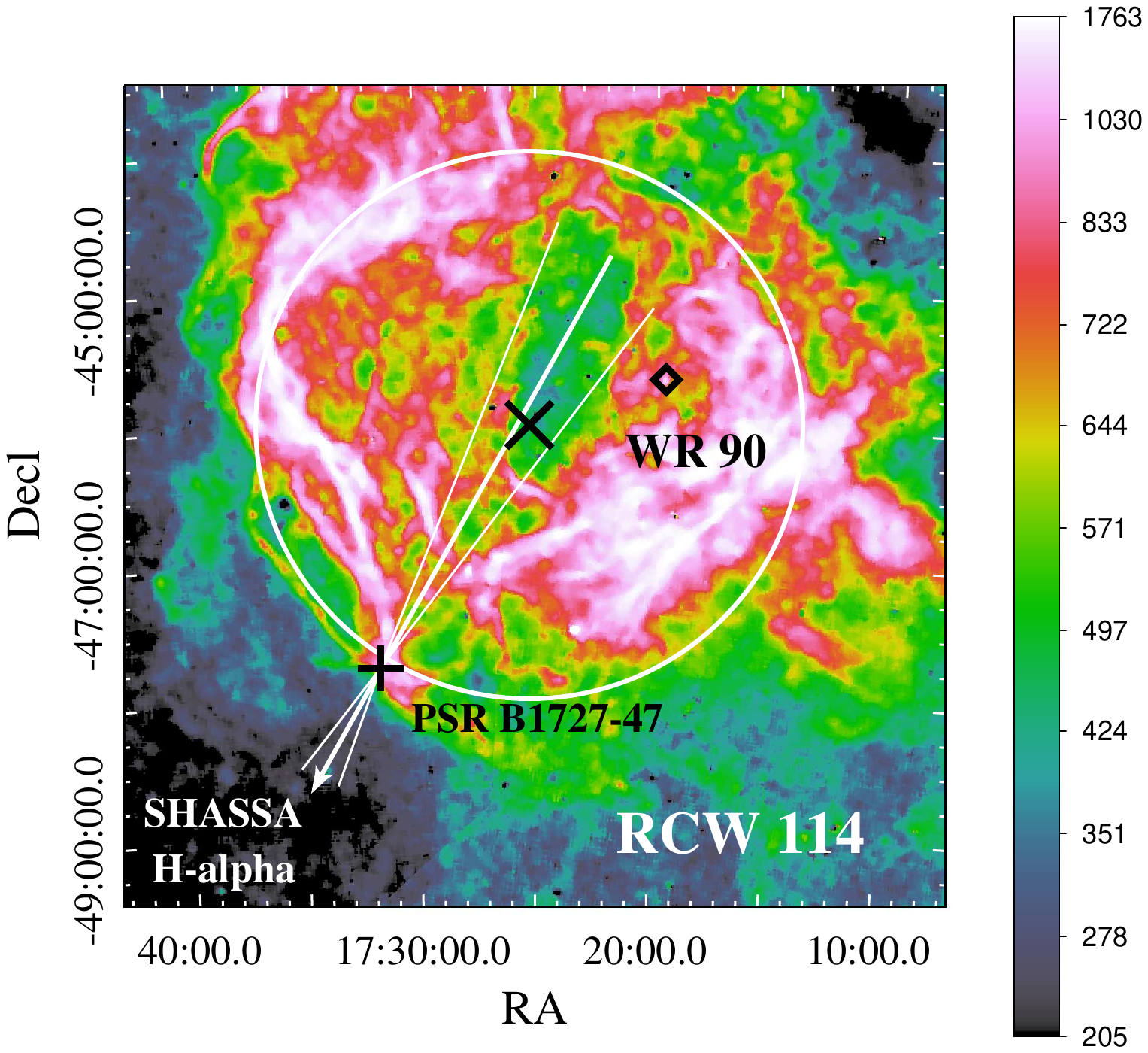}}
     \end{minipage}\hspace{0.05\textwidth}
     \begin{minipage}{0.45\textwidth}
     {\includegraphics[width=\textwidth,trim={3.3cm 6.5cm 2cm 6.5cm},clip]{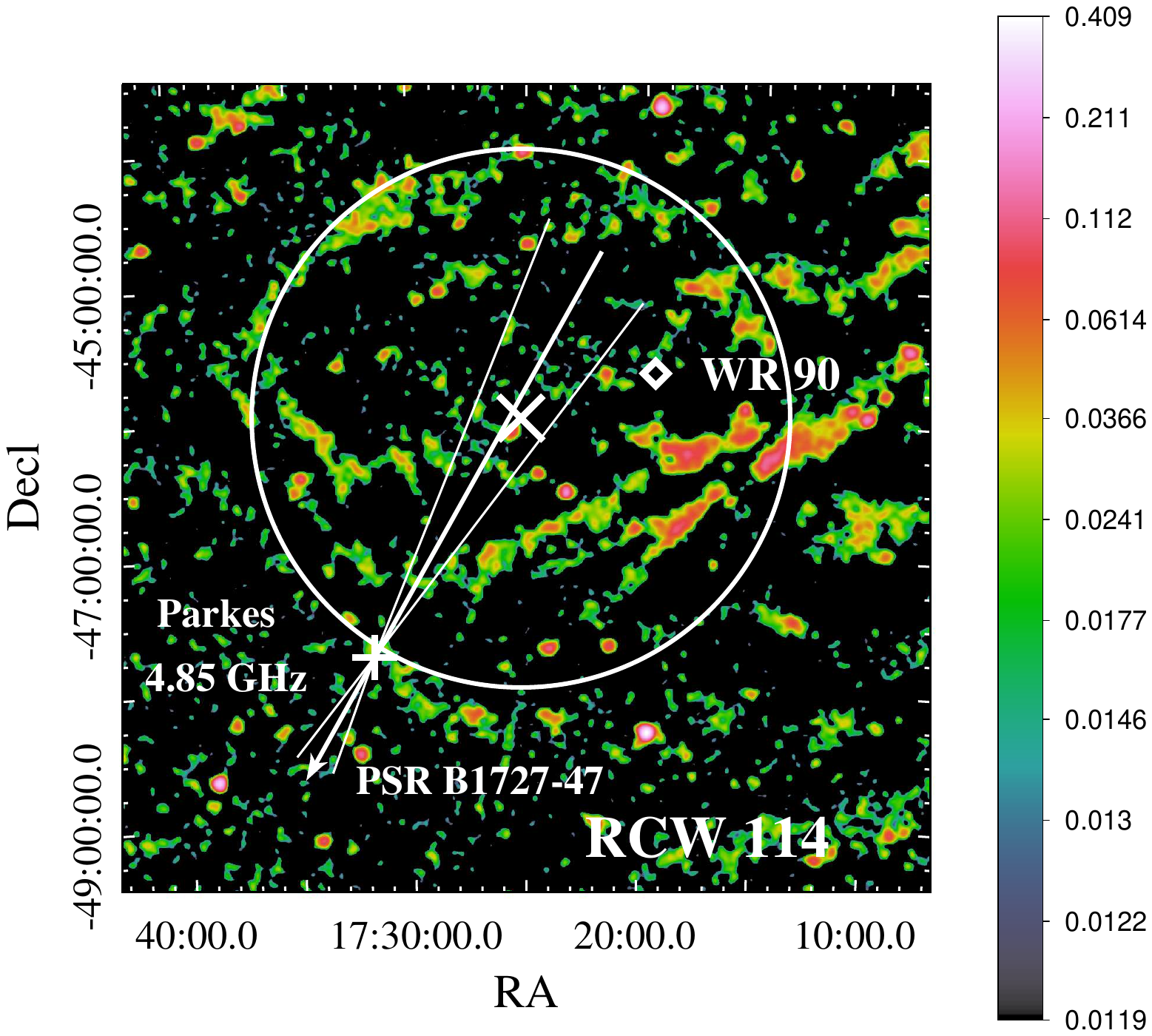}}
     \end{minipage}
     \begin{minipage}{0.45\textwidth}
     {\includegraphics[width=\textwidth,trim={3.3cm 6.5cm 2cm 6.5cm},clip]{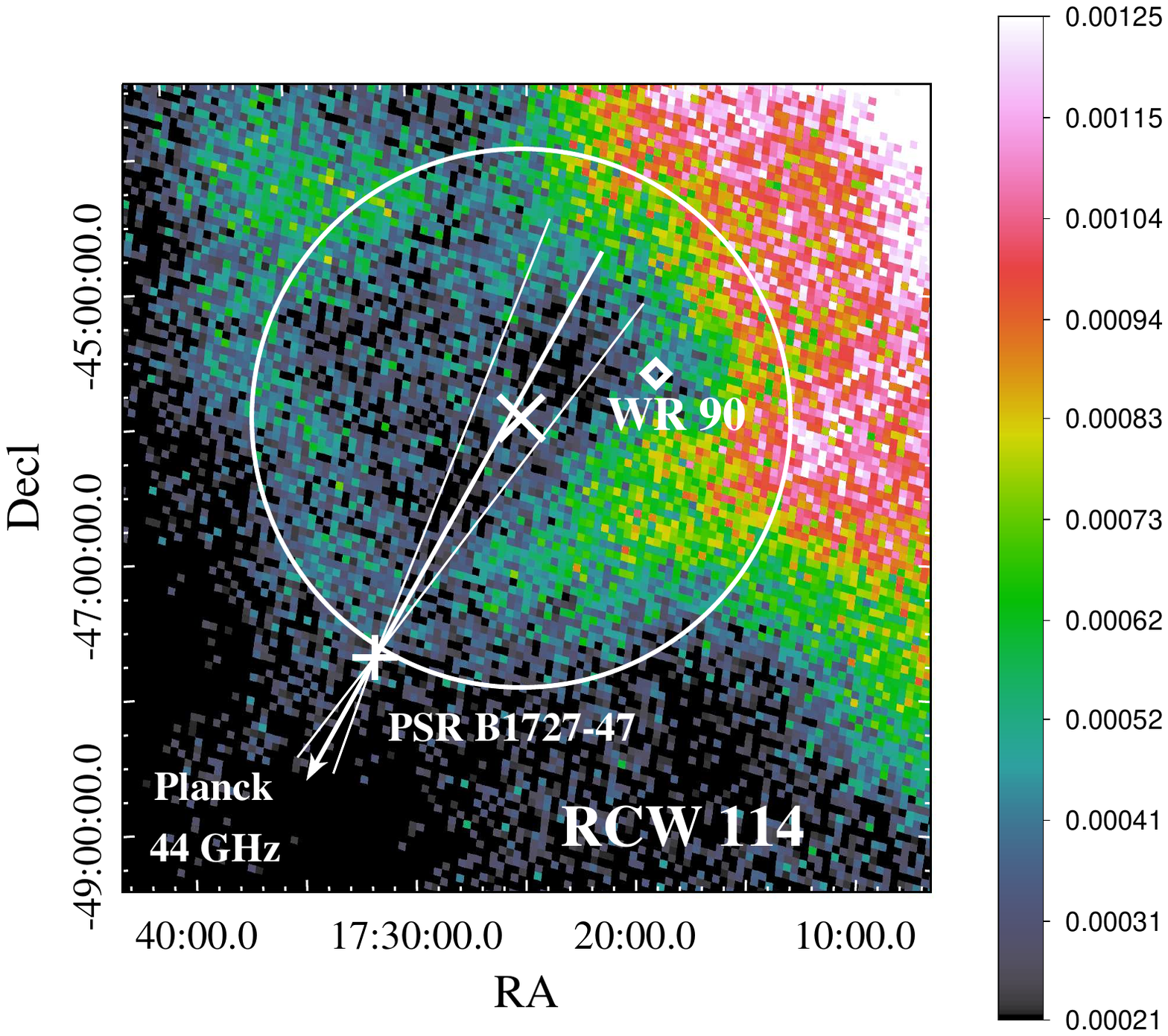}}
     \end{minipage}\hspace{0.05\textwidth}
     \begin{minipage}{0.45\textwidth}
     {\includegraphics[width=\textwidth,trim={3.3cm 6.5cm 2cm 6.5cm},clip]{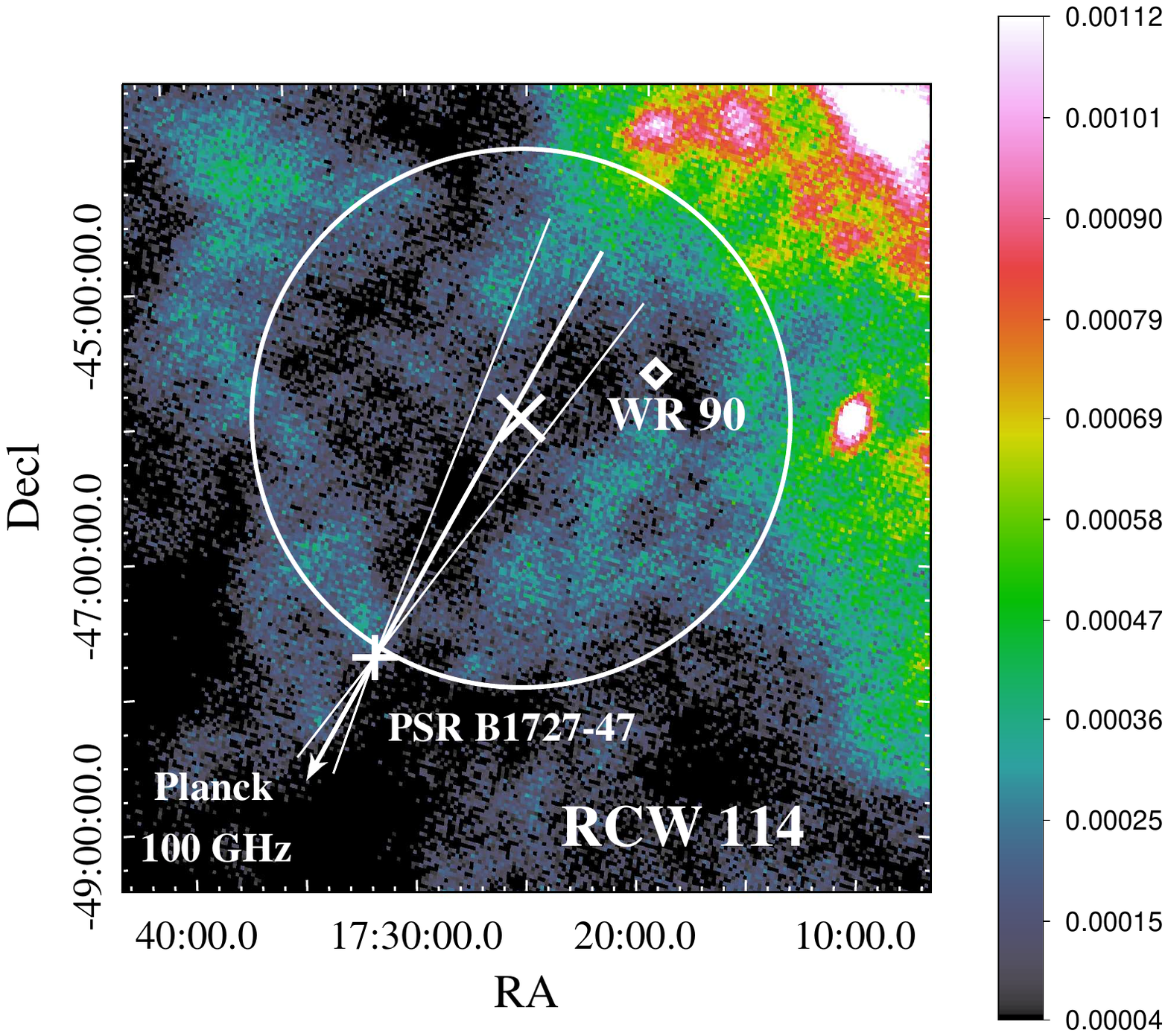}}
     \end{minipage}
   \end{center}
\caption{
RCW~114 SNR  images obtained in the H$\alpha$ emission line \protect\citep[The Southern H$\alpha$ Sky Survey Atlas (SHASSA);][]{2001gaustad},
at 4.85 GHz \protect\citep[The Parkes-MIT-NRAO Southern Sky Survey;][]{1993condon}, and 44 and 100 GHz (\textit{Planck} Release 1 results) as notified in the frames.
Colours show the intensity in  photons cm$^{-2}$ s$^{-1}$ sr$^{-1}$, in Jy beam$^{-1}$, and  in antenna temperatures (K), for the H$\alpha$, 4.85 GHz, and Planck images, respectively.
In each frame, the  circle of a 2\grad~radius roughly outlines
 the presumed outer shell of the SNR as seen in H$\alpha$;  the X-point is the circle center.  The diamond marks the position
 of the WR 90 star projected on the remnant.
 The B1727 position  is shown by the cross. Its p.m.~track  is presented by the  arrow, which
 is extended
 backwards
 by the characteristic age (80~kyr);
 thin lines show 1$\sigma$ uncertainties of the track. }
 \label{fig:RCW114}
\end{figure*}

\section{Association with RCW~114 }
\label{S:RCW114_assoc}
In the top left panel of Figure~\ref{fig:RCW114}, we show the H$\alpha$ image of the extended RCW~114 nebula whose shell structure is thought to represent an evolved SNR (see discussion in Section~\ref{S:RCW114} below). Three other panels of Figure~\ref{fig:RCW114} show radio images of  RCW~114 at 4.85 GHz (top right panel), 44 GHz (bottom left panel), and 100 GHz (bottom right panel)\footnote{The images were retrieved from the Internet Virtual Telescope archive using the SkyView tool: \url{https://skyview.gsfc.nasa.gov/current.}}.  In H$\alpha$, the shell-like  structure  of the nebulae with a radius of  $\approx$ 2\grad\ is clearly seen. In addition, three  almost parallel  ridges cross over the south-west SNR quadrant. Faint radio filaments spatially correlated with the bright  H$\alpha$ shell and the ridge features can be resolved in the 4.85 GHz radio map.
Although they are  only barely resolved at lower
frequencies  \citep{Duncan1997MNRAS},
the overall  shell-like structure of RCW~114 is
visible at higher frequencies  until  100 GHz in  the \textit{Planck} telescope data (bottom panels in Figure~\ref{fig:RCW114}).

B1727 position coincides with the south-east shell of this remnant, as shown by the plus sign in Figure~\ref{fig:RCW114}.
To check for the possible association between the pulsar and the SNR, we performed a backward extension of the B1727
p.m.~track, using the final p.m. vector measured in Section~\ref{S:comb} and the pulsar characteristic age of 80~kyr.
We checked that the pulsar track distortion
due to the Galactic gravitational potential is negligible at such a short time scale. This extension is shown by the
arrow in Figure~\ref{fig:RCW114}, while thin lines bracket the 68\% extrapolation uncertainty propagated from the measured p.m. uncertainty. The  pulsar track passes remarkably close to the putative RCW~114 shell center marked with the `x'-point in Figure~\ref{fig:RCW114} and shows that  B1727 was at least  born at a position within the remnant. Moreover, if the pulsar was  born at the remnant center, its age should be about  50~kyr, which is a reasonable value provided that  pulsar characteristic ages usually differ
from their true ages, when the latter are known \citep{Thorsett2003ApJ,PopovTurolla2012Ap&SS}. Difference between the inferred kinematic age and the pulsar characteristic age allows to constrain its initial period. We save this discussion to Section~\ref{S:birth}.

Thus we can conclude that the p.m.\ measurements of B1727 strongly support the genuine association between the PSR and the SNR. In this case the most plausible age of the PSR+SNR system is about 50~kyr.

\section{Discussion}\label{S:Discussions}

\subsection{Nature of the RCW~114 nebula}\label{S:RCW114}

For a long time the SNR nature of RCW~114 was not
obvious.
Historically, one of the counterarguments was  the absence of
the radio and soft X-ray emission typical for SNRs \citep{Bedford1984MNRAS}. However, Parkes and \textit{Planck} data shown in Figure~\ref{fig:RCW114} reveal faint radio emission spatially correlated with the nebula optical features.
In X-rays, RCW~114 has not yet been detected. Nevertheless, in the far-ultraviolet, the emission in the \CIV\ line spatially correlated with the H$\alpha$ features was found \citep{Kim2010ApJ}. This situation is possible
in evolved SNRs where the excited matter has already cooled down enough to be undetectable
in  X-rays while it can still be visible  in the ultraviolet \citep{1999Shelton}. Moreover, spectroscopic studies of the optical filaments in RCW~114 \citep{Walker2001MNRAS} revealed high \SII/H$\alpha$ line ratios, which are consistent with those produced by a supernova shock  in  interstellar matter \citep[ISM;][]{Fesen1985ApJ}.

According to the alternative interpretation \citep[e.g,][]{Cappa1988AJ,Welsh2003},
RCW~114  could be  an asymmetric bubble blown in ISM with a density gradient
by the wind from a
Wolf-Rayet  star HD~156385, also known as WR~90.  In  Figure~\ref{fig:RCW114}, the WR~90 location is marked  by the diamond.
However, analyzing \HI\ 21 cm maps that had better resolution than those used in the older studies, \citet{Kim2010ApJ} found two \HI\ voids in the direction of RCW~114. The prominent \HI\ void
already found by \citet{Cappa1988AJ} has a systemic velocity
of $\approx -4$~km~s$^{-1}$, whereas the velocity of a smaller  circular-shaped
void around  WR~90 is $\approx -13$~km~s$^{-1}$.
The latter structure  of about 50\amin~ in diameter is apparently visible
in  the H$\alpha$ and   \textit{Planck}  images in Figure~\ref{fig:RCW114} as well.
The prominent void can be identified with RCW~114, while the smaller void corresponds to the WR~90 wind-blown bubble. The difference in  velocities  suggests that RCW~114 is unrelated to WR~90 and, in addition, is closer to us \citep{Kim2010ApJ}.

Combining all the facts, we can confidently consider RCW 114  as an evolved SNR. Clearly, the presence of the neutron star, B1721, kicked roughly from the RCW~114 center, also favors the SNR interpretation.

\begin{figure}[t]
\setlength{\unitlength}{1mm}
     \includegraphics[scale=0.31, angle=0, clip]{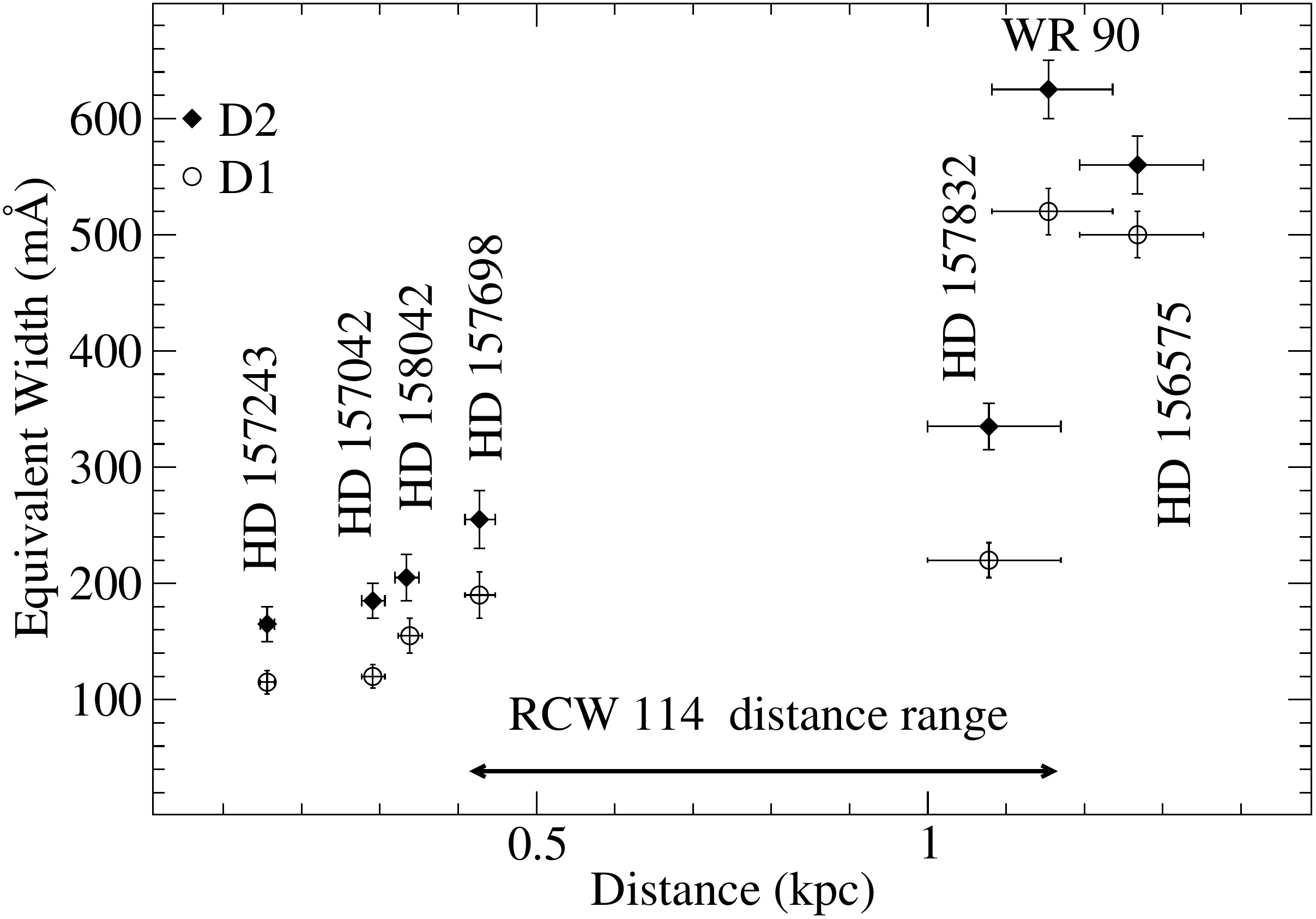}
  \caption{Interstellar \NaI\ D1 \& D2 absorption line equivalent widths {\sl vs} parallax-based distances for  seven stars projected on  RCW 114.
  The star names are shown near their data-points, the line widths are taken from \cite{Welsh2003}, and
  the parallaxes are measured
  with  \textit{Gaia}.
    The width excesses for WR 90 and HD 156575 are caused by the interstellar absorption associated with RCW 114 and the WR~90 \HI\ bubble.
  A plausible distance range for RCW 114 is shown by the double-ended arrow.
  }
  \label{fig:7stars}
\end{figure}

It is possible to get an estimate on the distance to the RCW~114 analyzing the spectral properties of the stars projected on the remnant. This was done by \citet{Welsh2003} who observed the interstellar \NaI\ D1 \& D2 Fraunhofer absorption lines  in spectra of seven field stars located within the distance range of 0.2--1.5 kpc.
We updated their analysis using  recent \textit{Gaia}
parallax-based distance measurements for these stars \citep{2018bailer-jones}.
In Figure~\ref{fig:7stars} we show the D1 \& D2 equivalent widths {\it vs.}
stars' distances based on the figure 3 in \citet{Welsh2003}. The prominent equivalent width excess at $D\ga 1.1$~kpc is much better localized as compared to the initial version  by \citet{Welsh2003}. Four nearby stars with $D\la 0.5$~kpc, including HD~157698 with a distance of
$0.43^{+0.020}_{-0.018}$~kpc, show a single component absorption profile indicating that they are foreground objects for RCW~114. The next by the distance star, HD~157832  ($D=1.078^{+0.092}_{-0.079}$ kpc), shows a broader double component profile demonstrating that it is already  behind (or within) the SNR. The WR~90 distance derived from recent \textit{Gaia} parallax measurements
is $1.154^{+0.082}_{-0.072}$ kpc \citep{2018bailer-jones}. This star, and HD~156575 ($D=1.268^{+0.084}_{-0.074}$ kpc), show even more complex  multi-component  structures of line profiles, with larger equivalent widths. A  similar pattern was found in  the IUE spectra of the interstellar
\SiII\ $\lambda1304$ absorption line profiles \citep{Welsh2003}.
This can be interpreted as a result of the contribution from both the RCW~114 filaments and the expanding WR~90 wind bubble \citep{Kim2010ApJ}. Notice that both WR~90 and HD~156575 project on the smaller \HI\ void \citep{Kim2010ApJ}, while HD~157832 does not.
The complexity of the profiles is  thus consistent with the \HI\ data also suggesting  that  WR~90 wind bubble is further than RCW~114 \citep{Kim2010ApJ}.
The spectroscopic analysis results in the 0.4--1.1~kpc distance range for the RCW~114 which is shown in Figure~\ref{fig:7stars} by the double-ended arrow.
This distance range is in accord with the limit of $\la 1.5$~kpc based on the assumption that the \CIV\ luminosity can hardly  exceed the luminosity of  Cygnus Loop \citep{Kim2010ApJ}. Observations of interstellar absorption features in spectra of field stars with the distances within the range of 0.5--1 kpc, filling the distance gap in Figure~\ref{fig:7stars} between HD 157698 and HD 157832, can help to better constrain the distance to the SNR.

\subsection{ RCW~114  {\sl vs} SNR  models, its ambient ISM density and   evolution phase }
\begin{figure*}[t!]
\begin{center}
\includegraphics[width=0.6\textwidth, clip]{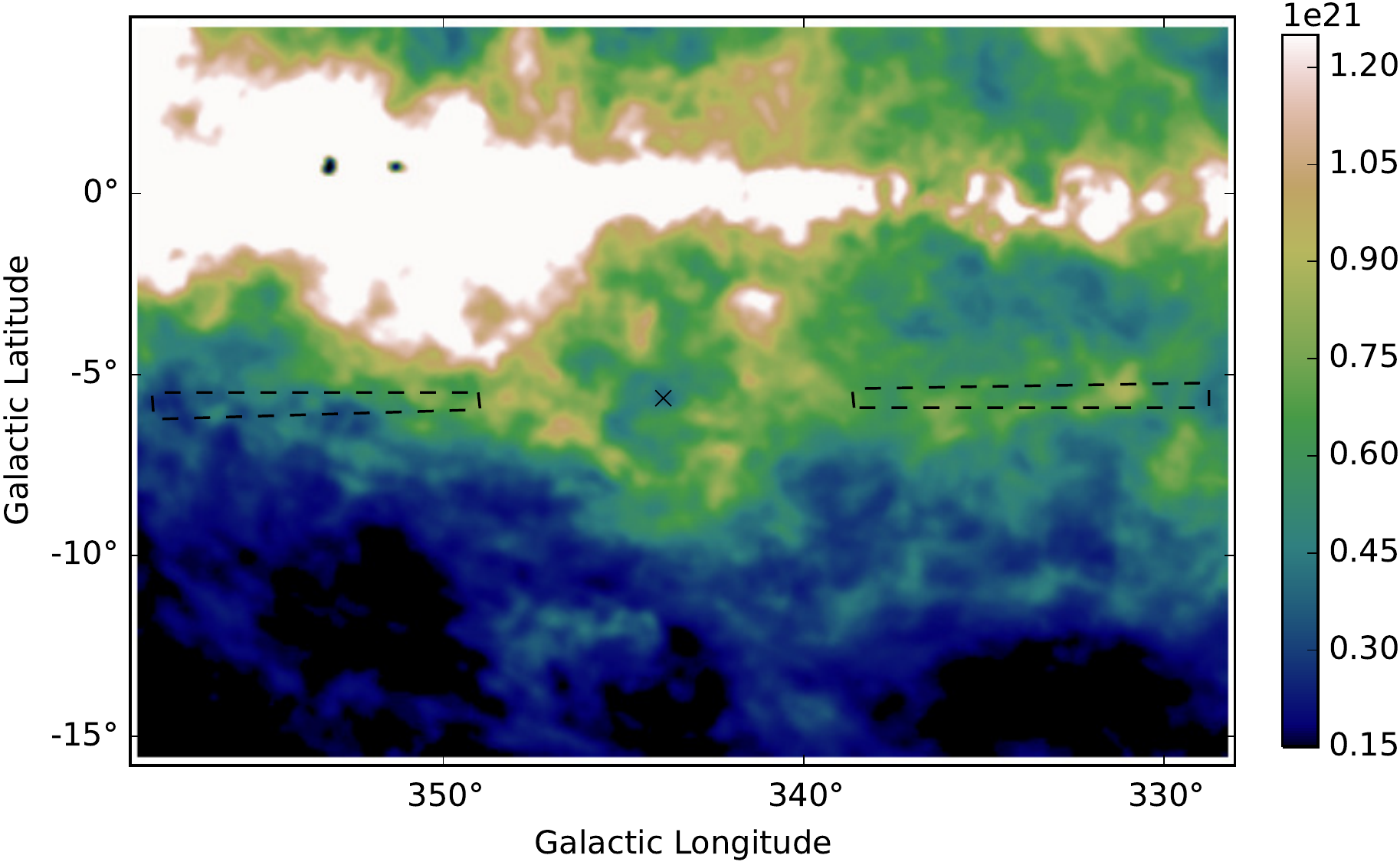}
\includegraphics[width=0.6\textwidth, clip]{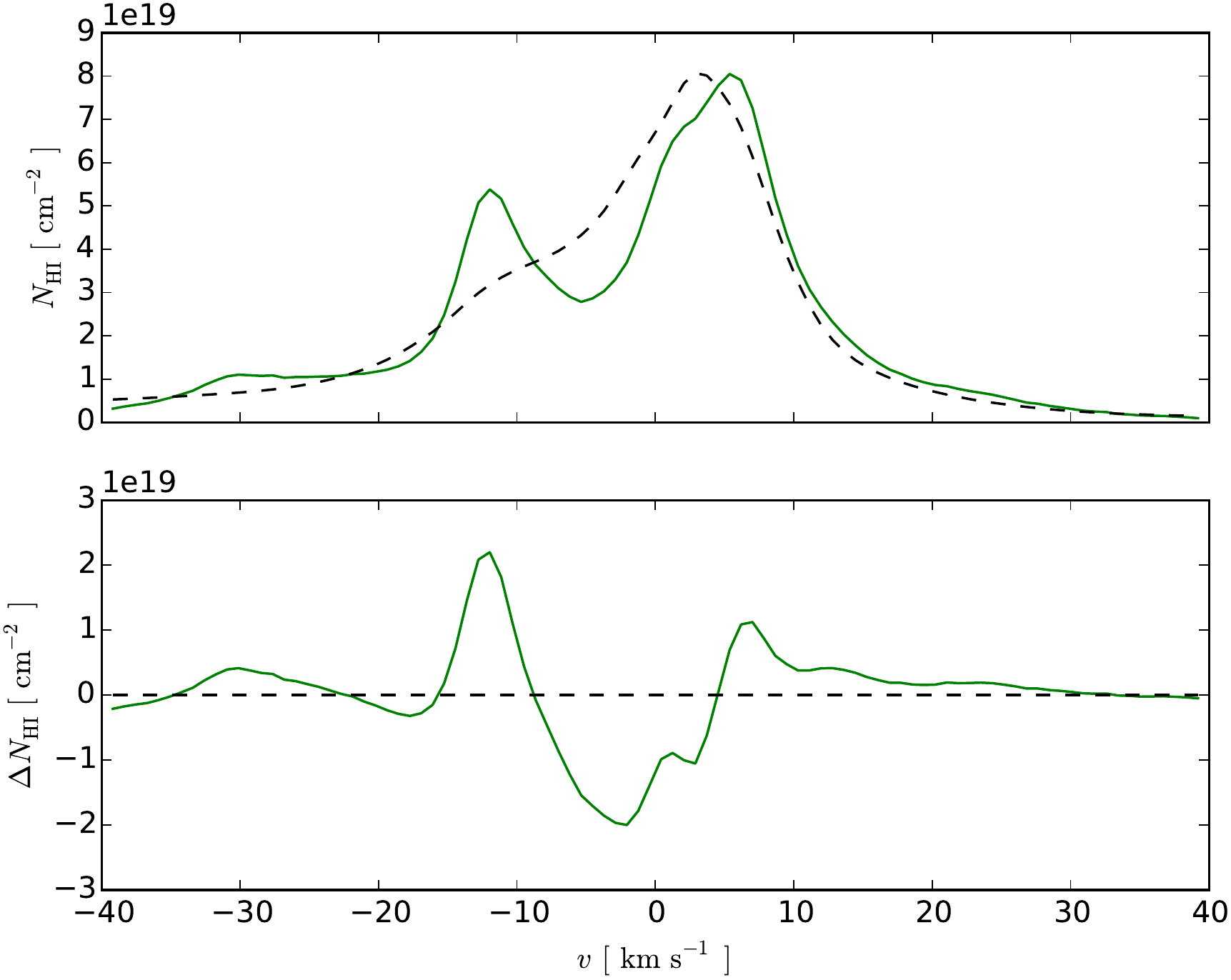}
\end{center}
  \caption{{ \sl Top:} GASS map of the \HI\ emission towards the RCW~114 SNR
  in units of $N_{\textrm{H}\,\textsc{i}}$ marginalized over the LSR velocities $-8\,
  {\rm km \, s^{-1}} < v < + 1 \, {\rm km \, s^{-1}}$.  RCW 114 is associated with the void whose approximate center is
  marked by the cross. Dashed polygons show  regions selected for the background extraction.
  { \sl Middle:}  The solid and dashed lines show the column density velocity profiles extracted
  along the direction marked by the cross   and the background regions in the top panel, respectively.
  { \sl Bottom:}  The difference  of the two curves
  in the middle panel  which reveals two
  ISM peaks
  and a void between them produced by the SNR shock.
  }
  \label{fig:hI_map}
\end{figure*}

It is important to check if the  pulsar-SNR association is consistent with  the observed RCW~114 properties.
The distance range of $0.4-1.1$~kpc and the angular diameter of 4\grad~imply the remnant radius in the range of
$16-35$~pc. Proper motion arguments suggest that the age of the system is about $50$~kyr (Section~\ref{S:comb}).
Both values are above the upper limits of $\approx$14~pc and $\approx$13~kyr constraining
 the Sedov-Taylor SNR expansion phase \citep[e.g.,][]{Cioffi1988ApJ}.
 Therefore the remnant has likely entered the next pressure-driven snowplough
 (PDS) phase.

  The analytical  expressions for the PDS expansion phase in a uniform ambient ISM with Solar   abundances from \citet{Cioffi1988ApJ}  result in the following
 approximate estimates for the SNR blast-wave shock velocity $V_{\rm s}$ and radius $R_{\rm s}$
\begin{eqnarray}
  V_{\rm s}&\approx& 134~{\rm km}~{\rm s}^{-1}\times E_{51}^{0.22} n_0^{-0.26} \left(\frac{t}{50~{\rm kyr}}\right)^{-0.7},\nonumber\\&&\label{eq:vpds}\\
  R_s&\approx& 22~{\rm pc} \times E_{51}^{0.22} n_0^{-0.26} \left(\frac{t}{50~{\rm kyr}}\right)^{0.3},\label{eq:rpds}
\end{eqnarray}
where $E_{51}$ is the supernova explosion energy in $10^{51}$ erg units,
$n_0$ is the pre-supernova  particle number density of the ambient ISM
in cm$^{-3}$,
and $t$ is the SNR age.
At $t\ga 20$ kyr, Equations~(\ref{eq:vpds})--(\ref{eq:rpds}) are consistent with more complex expressions by  \citet{Cioffi1988ApJ} within a few
per cent accuracy.  For the the remnant angular radius of  about  2\grad, Equation~(\ref{eq:rpds}) implies the distance
\begin{eqnarray}
  D&\approx& 647~{\rm pc} \times E_{51}^{0.22} n_0^{-0.26} \left(\frac{t}{50~{\rm kyr}}\right)^{0.3}\label{eq:Dpds} .
\end{eqnarray}
For reasonable values of $E_{51}$, $n_0$, and $t$, it is in accord with the 0.4--1.1~kpc range obtained in Section~\ref{S:RCW114}.

In order to further analyze Equations~(\ref{eq:vpds})--(\ref{eq:Dpds}), we tried to independently estimate $n_0$
using the third data
release of the Parkes Galactic All Sky Survey (GASS)
 of the Milky Way \HI\ emission \citep{Kalberla2015AsAp}.
The top panel in Figure~\ref{fig:hI_map} shows the \HI\ map of
the RCW 114  region summed up over the the local standard of rest (LSR) velocity range
$-8$~km~s$^{-1}$ $< v <$ $+1$~km~s$^{-1}$. The map is converted to the column density $N_{\textrm{H}\,\textsc{i}}$
using the expression $N_{\textrm{H}\,\textsc{i}} = 1.822\times
 10^{18} \Delta v T(v)$~cm$^{-2}$,
 where $T(v)$ is the brightness temperature in Kelvins  and  $\Delta
 v$ is the velocity channel width in km~s$^{-1}$ \citep[see, e.g.,][]{Matthews1998ApJ}.
 An extended \HI\ void is seen in the map which coincides by its position and
 morphology with RCW~114 (cf. Figure~\ref{fig:RCW114}), as discussed in Section~\ref{S:RCW114}. It presumably corresponds to a cavity in the ISM swept up by the passage of the SNR shock. An approximate void center is marked by the cross.

 The $N_{\textrm{H}\,\textsc{i}}$ velocity profile extracted from the data using the direction towards
 the center is shown in the middle panel of Figure~\ref{fig:hI_map} by the solid line.
  There is a dip in the profile between $-$10 km~s$^{-1}$ and 5
  km~s$^{-1}$ excavated  by the shock. The corresponding deficit of $N_{\textrm{H}\,\textsc{i}}$ divided by the
  SNR extent along the  direction would give us an estimate of the
  initial density of the ISM where the supernova exploded in. However, to estimate
  the deficit correctly,  the background gas
  has to be taken into account.
  To estimate the background, we take an average of $N_{\textrm{H}\,\textsc{i}}$ from two dashed polygons located outside  the SNR and stretched along the Galactic longitude (see the top panel of Figure~\ref{fig:hI_map}).
  The background velocity profile is shown by the dashed line
  in the middle panel of Figure~\ref{fig:hI_map}, while the background-subtracted  profile towards the SNR center  is presented in the bottom
  panel of Figure~\ref{fig:hI_map}.
  Two peaks or walls which are presumably formed by the material swept up
  by the SNR shock and a void between them are seen in the subtracted profile.
  The total $N_{\textrm{H}\,\textsc{i}}$ deficit between
  the two walls is $\approx 1.9 \times 10^{20} \, \rm cm^{-2}$.
The ambient matter density can be now estimated as $n_0=N_{\textrm{H}\,\textsc{i}}/(2R_s)$.
Notice that $R_s$ in Equation~(\ref{eq:rpds}) itself depends on $n_0$.

\begin{figure}[t!]
\includegraphics[width=0.95\columnwidth]{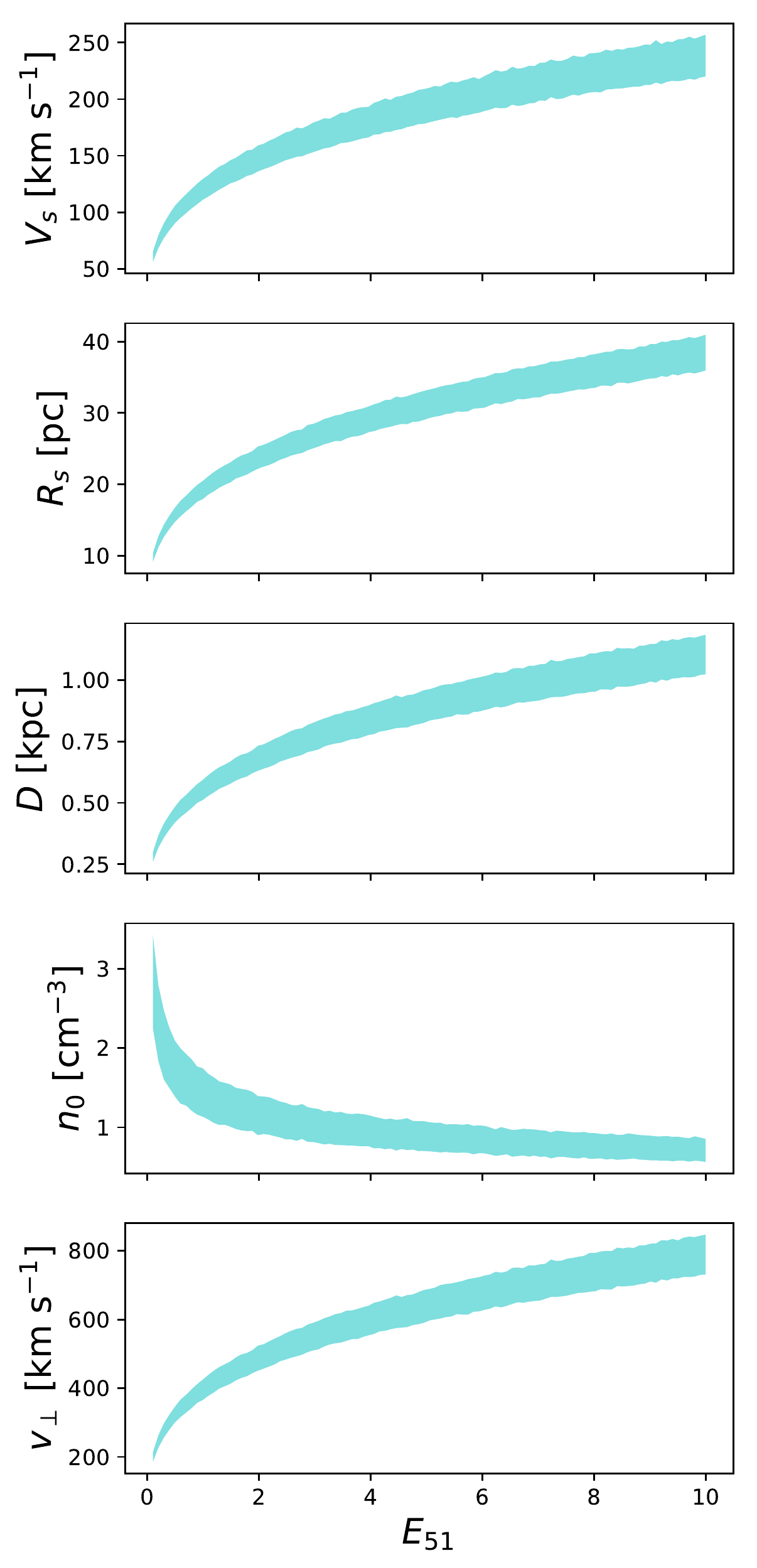}
\caption{Dependencies of the RCW~114 SNR parameters on the supernova explosion energy $E_{51}$. Panels show (from top to bottom) the expansion velocity $V_s$, the remnant radius $R_s$, the distance to the system $D$, the ambient matter density $n_0$, and the pulsar transverse velocity $v_\perp$ at the remnant distance. See text for details. }\label{fig:SNR_prop}
\end{figure}

This estimate allows us to exclude $n_0$ from Eqs.~(\ref{eq:vpds})--(\ref{eq:Dpds}), and the remnant properties are then parameterized only by the explosion energy $E_{51}$. In Figure~\ref{fig:SNR_prop} we show the dependencies of the parameters $V_s$, $R_s$, $D$, and $n_0$ on $E_{51}$ taking into account uncertainties in various constraints on parameters employed. The bottom panel in Figure~\ref{fig:SNR_prop} additionally shows the pulsar transverse velocity $v_\perp$ at the RCW~114 distance. Specifically, to obtain the uncertainty strips plotted in Figure~\ref{fig:SNR_prop}, we included $0\fdg 1$ uncertainty on the remnant angular radius, $0\fdg 3$ uncertainty on the pulsar shift during lifetime (this, combined with the B1727 p.m., gives the system age, cf. Section~\ref{S:RCW114_assoc}), and 20\% accuracy of our $N_{\textrm{H}\,\textsc{i}}$ deficit estimate. As can be seen from Figure~\ref{fig:SNR_prop}, our adopted distance range of $0.4-1.1$~kpc corresponds to a broad explosion energy range of $(0.3-10)\times 10^{51}$~erg.

Typical explosion energies obtained in simulations of neutrino-driven core-collapse supernovae are about $(0.5-1.5)\times 10^{51}$~erg \citep{Sukhbold2016ApJ,Muller2016MNRAS}. This puts RCW~114 at the lower edge of
the adopted  distance range. For instance, for $E_{51}=1$ the dependencies shown in Figure~\ref{fig:SNR_prop} result in $V_s=120\pm10$~km~s$^{-1}$, $R_s=19\pm 1$~pc, $D=550\pm 40$~pc, $n_0=1.4\pm0.3$~cm$^{-3}$, and $v_\perp=397\pm 30$~km~s$^{-1}$.
Most of the  simulations  suggest
an explosion energy upper limit   of about $2\times 10^{51}$~erg for of the core-collapse neutrino-driven supernovae \citep[][and references therein]{Janka2012ARNPS,Muller2017IAUS} \citep[see, however,][]{Pejcha2015ApJ}. As follows from Figure~\ref{fig:SNR_prop}, at $D\ga 750$~pc the RCW~114  explosion energy should be higher than this limit, which in turn means that a non-standard 
explosion mechanism, e.g., a  magneto-rotational supernova, needs to be invoked in this case.
This suggests that the lower half of the adopted distance range, i.e.,
400--750~pc, is somewhat more probable. 

At the same time, the estimated shock speeds are $V_{\rm s}>95$~km~s$^{-1}$ (top panel in Figure~\ref{fig:SNR_prop}) which is hard to reconcile with the
expansion
velocities
of 25--35~km~s$^{-1}$ obtained in the analysis of the radial velocities measured in the optical for a few RCW~114 filaments  \citep{Meaburn1991}.
This is not unusual that the radial speeds measured in the remnant filaments are smaller than the real expansion speed.
In this case,
the lower-speed motion of the filaments may be driven by a secondary remnant shock. A similar situation is observed, e.g., for another evolved SNR S147 associated with PSR J0538$+$2817. In this case, the maximal measured velocity
is about 100~km~s$^{-1}$ while most filaments show significantly smaller velocities, with a mean value of 10~km~s$^{-1}$ \citep[e.g.,][and references therein]{Ren2018}.
Moreover, for one faint  filament located at the south-east edge of the RCW~114 remnant,  \citet{Meaburn1991}  found  a high velocity H$\alpha$
component centered at 80~km~s$^{-1}$ with  the line width of about 30~km~s$^{-1}$. If it is associated with the remnant expansion, the real expansion speed should be higher due to the projection arguments. This is in agreement with
the calculated $V_{\rm s}$ range.

\subsection{Pulsar birth period}\label{S:birth}
The independent estimation of the pulsar age based on the derived p.m. and the PSR-SNR association allows one to probe its initial period. For the power-law spin-down evolution with a braking index $n$, the pulsar initial period can be found from the true age $\tau$ and the current period $P$ as \citep[e.g.,][]{Lorimer2012hpa}
\begin{equation}\label{eq:init_period}
P_0=P \left[1-\frac{n-1}{2} \frac{\tau}{\tau_c}\right]^{1/(n-1)}.
\end{equation}
The purely magnetic dipole braking mechanism corresponds to $n=3$.
Then, assuming $\tau=50\pm10$~kyr, we get $P_{0,3}=0.5\pm0.1$~s (here we follow the notation of \citet{Noutsos2013MNRAS} where the second number in the initial period superscript, if present, corresponds to the adopted value of $n$).
Braking indices are rarely known well. In cases when the braking indices are reliably measured, they are less than 3, making the simple dipole spin-down model inadequate. Since \b1727\ has a lot of glitches and timing noise, it is hardly possible to firmly estimate $n$ ($\ddot{\nu}$) from timing solution.
Allowing $n$ to vary in the range of $ 1.4-3$ we get $P_0=0.56\pm 0.07$~s,
where the error is dominated by the age uncertainty and not by the adopted index range.  Therefore, if the supernova explosion was highly asymmetric with
the pulsar birth site
significantly shifted from the SNR geometric center, \b1727\ could be older and thus faster rotating at birth.

The inferred initial period  of  \b1727\  is quite large as compared to the $P_0$ values
of  30 young pulsars whose  real ages were more or less reliably estimated via their associations with  SNRs   \citep{PopovTurolla2012Ap&SS}. With a few exceptions, their initial periods 
cluster  at $P_{0,3}\lesssim 0.1$~s.
\citet{Noutsos2013MNRAS} analyzed kinematic ages of 27 pulsars and  found a bimodal distribution of pulsars over $P_{0,3}$.
The long-period component formed
by a half-dozen of pulsars is located at
$P_{0,3}>0.6$~s.  \b1727\ could belong to this population.
However, its members  are significantly older, $>10^5$~yr, than \b1727\ as well as the sample presented by \cite{PopovTurolla2012Ap&SS}.
It is not impossible that the appearance of this population in the analysis by \citet{Noutsos2013MNRAS}
is related to the  pulsar magnetic field evolution and does not resemble the actual initial period distribution   \citep{IgoshevPopov2013MNRAS}.
\b1727\ is young and as such, the longer term evolution of the magnetic field can hardly affect  its initial period estimate.

There are only two objects in the \cite{PopovTurolla2012Ap&SS} sample that have $P_{0,3}>0.4$~s. One of them, PSR~J1210$-$5225 associated with the SNR G296.5$+$10, is a weakly-magnetized, $B\sim 10^{11}$, slowly rotating pulsar dubbed ``anti-magnetar'' \citep[e.g.,][and references therein]{Gotthelf2013ApJ,Halpern2015ApJ}. It is also known as  CCO 1E1207+5209 and is a peculiar object in comparison to a bulk of NS populations  \cite[see, e.g.,][for review]{Deluca2017JPhCS}. In contrast, the second source, PSR~B2334$+$61 associated with the SNR G114.3$+$0.3, is quite similar to \b1727. It has $B\approx10^{13}$~G, the period $P=0.49$~s, and a characteristic age of $40$~kyr. Similarly to \b1727, its association with the SNR suggests a lower distance than inferred from the electron density maps and a smaller true age $\tau \sim 10$~kyr \citep{YarUyaniker2004ApJ}. The latter translates to $P_{0,3}\approx 0.9 P=0.45$~s,
which is remarkably close to our estimate for \b1727. On the other hand, only a marginal p.m. of $\mu_\alpha=-1\pm 18$~mas~yr$^{-1}$ and $\mu_\delta=-15\pm16$~mas~yr$^{-1}$ is reported for PSR~B2334+61 \citep{hobbs2005MNRAS} resulting in a 3$\sigma$ upper limit of the transverse velocity  $v_\perp< 230 D_{0.7~\mathrm{kpc}}$~km~s$^{-1}$, where $D_{0.7~\mathrm{kpc}}$ is the distance to PSR~B2334$+$61 scaled by the favored distance of 0.7~kpc \citep{YarUyaniker2004ApJ}.
Although somewhat smaller, this value is nethertheless comparable to the transverse velocity inferred for \b1727.

Recent simulations of the core-collapse supernovae by \citet{Muller2019MNRAS} revealed a significant anti-correlation between NS $P_0$ and $v_\perp$. On the other hand, parametric simulations by \citet{Wongwathanarat2013A&A} resulted in  the initial periods  in the range of $0.1-8$~s and the birth kick velocities up to more than $700$~km~s$^{-1}$ (due to a so-called ``tug-boat'' mechanism) without any apparent correlation between the two parameters.
The relatively large initial periods accompanied by  the large kick velocities  inferred for \b1727,  and partially for PSR~B2334$+$61, appear to be challenging for both simulations, which make   these two pulsars important tests   for the core-collapse  supernova models.

\subsection{Pulsar distance and the Galactic electron density distribution}
Current models of the electron density distribution in the Galaxy place B1727 either at the DM distance of 2.7~kpc  \citep[NE2001 model;][]{cordes2002astro.ph} or even at 5.5~kpc
 \citep[more recent model by][]{yao2017}. According to our p.m.\ measurements (Section~\ref{S:comb}),  both models result in an unrealistically  high transverse velocity $v_{\perp}$  of about 2000 or 4000 km~s$^{-1}$, respectively. Typical pulsar transverse velocities are in the range of $50-700$~km~s$^{-1}$ \citep{brisken2003AJ,hobbs2005MNRAS,Faucher2006ApJ,Verbunt2017A&A,Jankowski2019MNRAS} suggesting that B1727 is much closer than the electron density  models predict. This is again in line with the picture of the association of B1727 with RCW~114, see the bottom panel in Figure~\ref{fig:SNR_prop}. The adopted  RCW~114 distance range   of 0.4--1.1~kpc results in the pulsar $v_{\perp}$ in the range of 300--800~km~s$^{-1}$, with the higher plausibility being granted to its lower end. This makes the pulsar velocity a typical one, supporting further the PSR-SNR association. As stated above, such a value of the pulsar velocity can be
  explained by the standard pulsar kick mechanisms \citep[e.g.,][and references therein]{Janka2017ApJ}.
In this case, however, the existing models of the Galactic electron density distribution   require serious corrections in the direction to the pulsar.

An independent support for the smaller distance comes from the
recent $\gamma$-ray detection of \b1727\ \citep{Smith2019}. The $\gamma$-ray efficiency $\eta=L_\gamma/\dot{E}$, i.e. the ratio of the $\gamma$-ray luminosity to the spin-down energy, is found to be 80\% for the distance of 2.7~kpc and more than the limiting 100\% for the 5.5~kpc (assuming the $\gamma$-ray beaming factor $f_\Omega=1$). In contrast, at the distances less than $1$~kpc, one obtains $\eta\lesssim 11\%$, the value typical for the $\gamma$-ray pulsars. Therefore, the $\gamma$-ray observations also point out to the overestimation of the \b1727\ distance by the current electron density models, although the possibility of a strong beaming ($f_\Omega \ll 1$) or an unknown contamination of the observed $\gamma$-ray flux by RCW~114 can somewhat lighten these arguments \citep{Smith2019}.

The
discussion above implies that the  fractional uncertainty of the models of the Galactic electron
density distribution along the pulsar line of sight is  60--80\%.
Such a situation is not unusual \citep{yao2017}. For instance,
the very long baseline interferometry (VLBI) parallax measurements for two   pulsars revealed
65\% lower distances than followed from the NE2001 model
\citep{Kirsten2015AA}. Subsequent extensive VLBI parallax analysis of 57 pulsar under the PSR$\pi$ project also suggests that the accuracy of the DM-based distance predictions is significantly overestimated \citep{Deller2018arXiv}.
Using precise timing observations of 42
millisecond pulsars, \citet{Desvignes2016MNRAS} found that the
fractional NE2001 uncertainty can be as high as 80\%.
 For the distance resulting from  the pulsar--RCW~114 association, the NE2001 model predicts a lower DM of B1727 than the actual one. This
indicates the presence of an  unmodelled nearby electron `clump' on
the line of sight.
For the model of \citet{yao2017},
such a clump should
be even denser or larger.

\subsection{Properties of B1727 glitches}\label{S:glitches}
In this work we found two new glitches of B1727.  Four events were reported previously \citep[see][and references therein]{Yu2013MNRAS} and
\citet{Jankowski2017ATel} recently reported on the newest and largest glitch of B1727  with $\Delta \nu_{\rm g}/\nu\approx3.148 \times10^{-6}$ which occurred in August 2017\footnote{The exact date of the glitch is unknown, since it happened during a gap in observations \citep{Jankowski2017ATel}.}. The seven glitches
 are distributed
over a 23.5 yr data span.
This suggests a mean glitch rate of $\sim$0.25 yr$^{-1}$.
 It is consistent with the observed
rate  for pulsars  with relatively high pulse frequency derivatives $\dot{\nu}$ between a few times of $10^{-13}$ and $10^{-11}$ ~Hz s$^{-1}$
\citep{2011espinoza}.
For B1727, $\dot{\nu}\approx2.37\times10^{-13}$~Hz s$^{-1}$ and is marginally within the above range.

\begin{figure}[ht]
\includegraphics[width=\columnwidth]{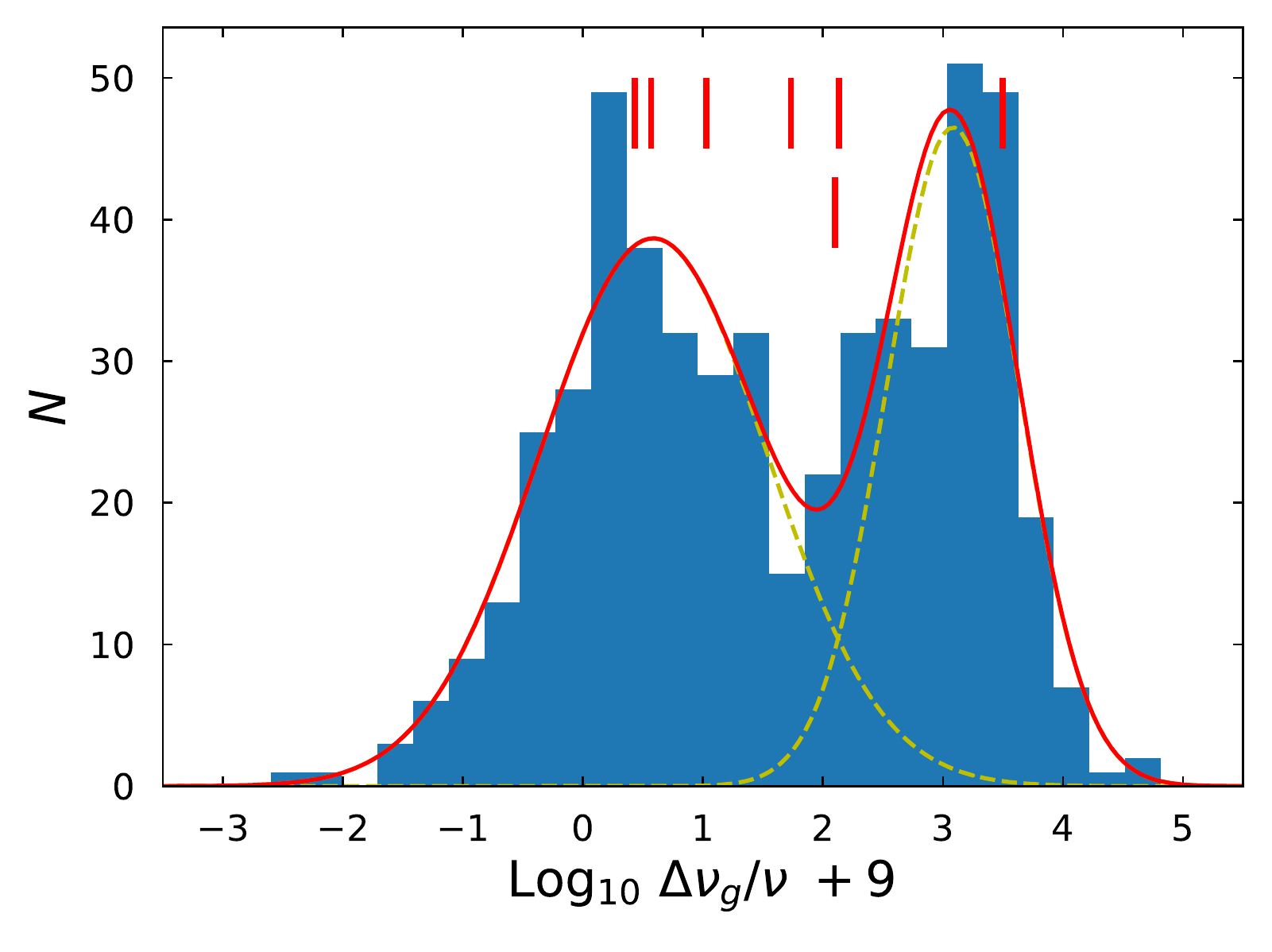}
\caption{Distribution of relative  pulsar glitch sizes according
to the data in \url{http://www.jb.man.ac.uk/pulsar/glitches.html},
as given for August 2018. The solid line is the double-Gaussian
best fit for the distribution; dashed lines show individual
components. Vertical bars mark the positions of seven glitches of
B1727 discussed in the text.}\label{fig:glitch_distribution}
\end{figure}

The fractional glitch size $\Delta \nu_{\rm g}/\nu$ for B1727 varies in the range of (2.7--3147.7)$\times 10^{-9}$.
 The data on $\Delta \nu_{\rm g}/\nu$ collected for all pulsars demonstrate a bimodal distribution with
 two  wide peaks \citep[][and references therein]{2011espinoza, Yu2013MNRAS}. This distribution is shown in Figure~\ref{fig:glitch_distribution} by the blue histogram, with the solid line showing the best-fit two-component Gaussian distribution. Individual Gaussian components are plotted with dashed lines in Figure~\ref{fig:glitch_distribution}.

 The small glitch peak is located around $\Delta \nu_g/\nu\approx 3\times10^{-9}$ and the large glitch peak  is near $10^{-6}$. The peaks are
 separated by a dip at $\Delta \nu_{\rm g}/\nu \sim 10^{-7}$. This distribution
 suggests that two different  mechanisms may trigger small and large glitches.
 The cracking of the NS crust may be responsible for the small glitches, while
 other mechanisms, such as a critical superfluid vortex repining inside the star
 (e.g., ``snowplough'' glitch model) need to be invoked to explain large glitches as observed, i.e., in the Vela pulsar \citep{Haskell2015IJMPD}.
 Fractional sizes of the B1727 glitches are shown with the vertical bars in Figure~\ref{fig:glitch_distribution}. As seen, they do not follow the overall distribution. Three glitches of B1727 with $\Delta \nu_{\rm g}/\nu\approx2.7,\,3.7,\, 10.7 \times10^{-9}$ are compatible with the small glitch population, the largest recently discovered one clearly belongs to the large glitch population, while the rest three  glitches with $\Delta \nu_{\rm g}/\nu\approx 54,\, 126,\, 136 \times10^{-9}$
fall exactly in the dip of the glitch distribution.
\begin{figure}[t!]
\includegraphics[width=\columnwidth]{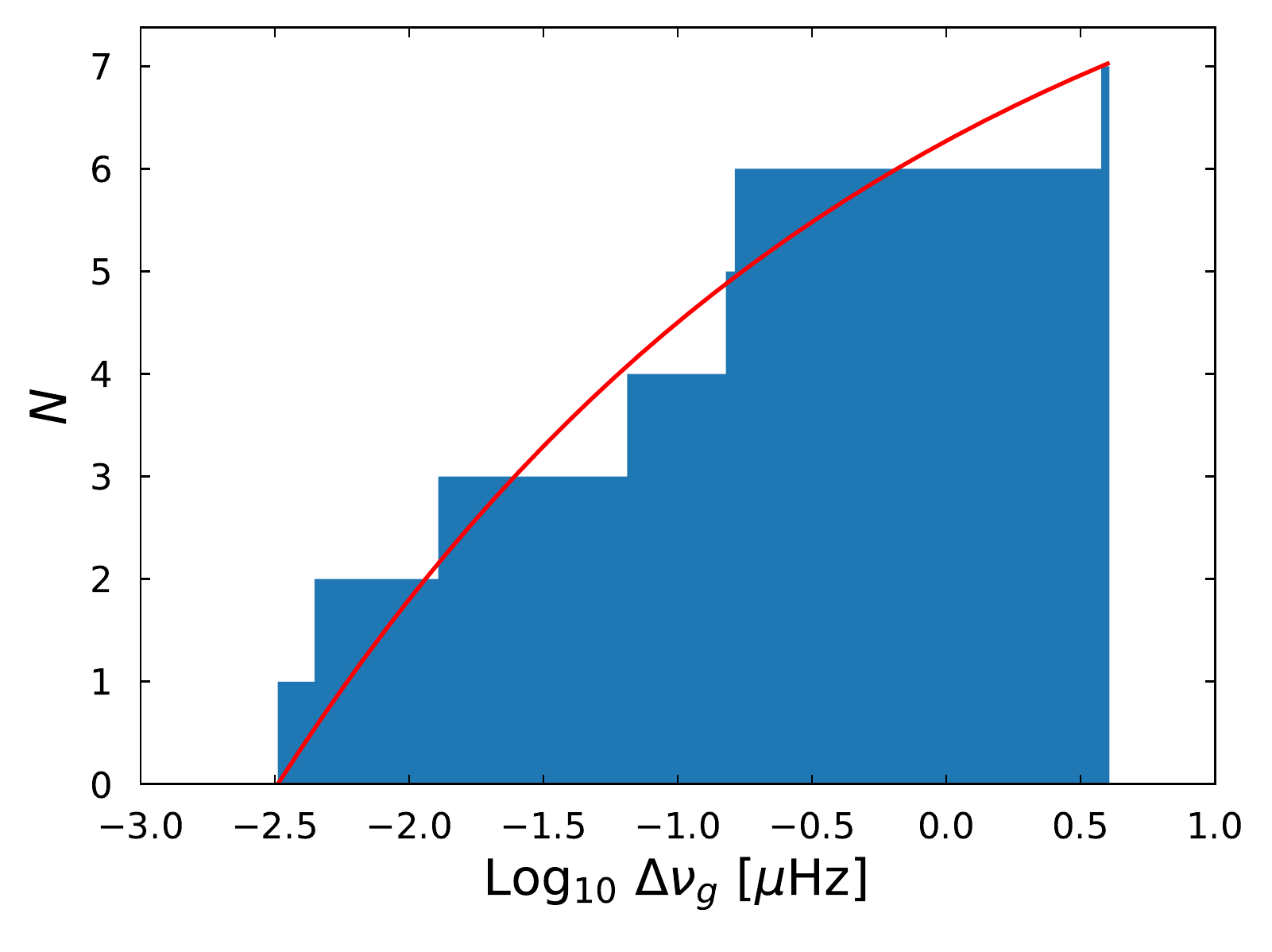}
\caption{Cumulative distribution of the glitch sizes observed in
B1727. The solid line shows the best-fit power law
distribution.}\label{fig:glitch_sizes}
\end{figure}
Seven glitches detected in B1727 allows to roughly investigate the glitch statistics for this pulsar. In Figure~\ref{fig:glitch_sizes}, we show the cumulative distribution of glitch sizes (i.e. the number of glitches with $\Delta \nu_g$ smaller than the given one). The obtained broad distribution is similar to those for the Crab and similar pulsars \citep{2011espinoza} in contrast to Vela-like glitchers. The distribution in Figure~\ref{fig:glitch_sizes} is well-fitted by the power-law \citep[see, e.g.,][]{Melatos2008ApJ} with the index $\alpha=1.2\pm0.3$ in agreement with the values found for other frequently glitching pulsars \citep{Melatos2008ApJ, Haskell2015IJMPD}. The best-fit power law is shown with the solid line in Figure~\ref{fig:glitch_sizes}. Figure~\ref{fig:glitch_sizes} indicates that probably the same trigger mechanism operates in B1727, including the recent largest event.
 The glitch spin-up rate $\Delta \dot{\nu_g}$ for B1727 does not vary substantially from glitch to glitch and
 is consistent
 with what is typically observed for other pulsars of similar ages \citep{2011espinoza}.

 The exponential and fractional glitch recovery parameters $\tau_d$ and $Q$
 were estimated only for three
 of seven B1727  glitches.  The $Q$ histogram for all pulsars  is also  bimodal,
 pointing on two different mechanisms
 of the glitch recovery
 \citep{Yu2013MNRAS}.
 The first peak is located near $Q\approx 0.01$ and the second is between 0.1 and 1. Interestingly, for all three glitches where $Q$ value was estimated, it falls in the dip between the two peaks of the distribution making it impossible to speculate about mechanisms of the glitch recoveries in B1727.

\section{Conclusions}\label{S:Conclusions}

We measured, for the first time, the p.m.~of the radio-bright
pulsar
B1727. Timing observations during 50 yr, since the pulsar discovery,
and the interferometric positions for three epochs resulted in significant  $\mu_{\alpha} = 73\pm 15$ mas~yr$^{-1}$ and $\mu_{\delta} = -132\pm 14$ mas~yr$^{-1}$.

The obtained p.m.~allowed us to
reveal the birth place of B1727 near the center of the nearby evolved SNR
RCW~114. This strongly suggests the genuine association between
the pulsar and the SNR. The association enabled us to  constrain self-consistently the properties of the system.

Based on the analysis of all the data, we argue  that the system
is located at $\approx 0.4-0.7$~kpc, much closer than given by the
pulsar dispersion measure and current Galactic electron density
maps. The dispersion measure distance results in unrealistically
high transverse velocities of $\ga 1900$~km~s$^{-1}$. In contrast,
the new distance and p.m.~imply a reasonable pulsar transverse
velocity of about 300--500~km~s$^{-1}$. The inferred distance
makes feasible B1727 parallax measurements with the VLBI which can
help to set stronger constraints on the system parameters and to
correct the electron density distribution towards the
pulsar\footnote{First sets of Long Baseline Array (LBA)
observations have been carried out in July, September and December
2018.}.

Based on the measured p.m. value and the pulsar shift from the
remnant center, we conclude that the most plausible system age is
about 50~kyr. For reasonable values of braking indices, this
implies a relatively large pulsar initial period of about
0.5$-$0.6~s.  Comparison of the RCW~114 data with the SNR
evolution models shows that the remnant is  at the middle of the
PDS phase. The remnant properties are consistent with being a
product of a usual neutrino-driven core-collapse supernovae. This
places  the system  to the lower edge of the above  distance
range.

We detected two new glitches of the pulsar. Together with four  glitches found previously and that of August 2017
this suggests a mean glitch rate for B1727 of 0.25 yr$^{-1}$. The pulsar likely demonstrates a broad range of the glitch sizes described by the power-law distribution.

\acknowledgments{
We are grateful to the anonymous referee for useful comments that helped us improve the paper.
We thank S.~A.~Balashev, G.~B.~Hobbs, L.~Lentati, R.~N.~Manchester, and D.~A.~Smith for
discussions.

{The Australia Telescope Compact Array is part of the Australia Telescope National Facility which
is funded by the Australian Government for operation as a National Facility managed by CSIRO. This paper includes archived
data obtained through the Australia Telescope Online Archive (\url{http://atoa.atnf.csiro.au)}.}}

\vspace{5mm}
\facilities{ATCA, Parkes, \textit{Planck}, \textit{Gaia}}

\software{MIRIAD \citep{miriad}, Karma \citep{gooch}, TEMPONEST \citep{Lentati2014MNRAS}, PSRCHIVE \citep{Hotan2004}, TEMPO2 \citep{Hobbs2006MNRASTempo2,Edwards2006MNRASTempo2}, AstroPy \citep{Astropy2013A&A,Astropy2018AJ},  SciPy \citep{Scipy}, Matplotlib \citep{Hunter:2007}.}

\appendix
\section{ATCA relative astrometry}\label{S:atca_rel}
\begin{deluxetable*}{*{8}{l}}[ht]
\tablecaption{Reference sources used for the registration of the ATCA images.\label{tab:t-atca-ref-sour-samesc}}
\tablehead{
\colhead{Source \#}  &  \colhead{R.A., J2000} & \colhead{Decl., J2000} & \colhead{$S/N$}  & \colhead{$\chi^2_{11-16}$} & \colhead{$\chi^2_{11-16}$}  & \colhead{$\chi^2_{11-05}$} & \colhead{$\chi^2_{11-05}$}
  \\
  \colhead{}& \colhead{} & \colhead{}  & \colhead{}  & \colhead{before} & \colhead{after}& \colhead{before} & \colhead{after}
}
\startdata
1 & 17\h31\m01\fss385(15) & $-$47\degs29\amin44\farcs37(22)  & 14  & 55  & 0.41& 8.6 & 0.9 \\
2 & 17\h30\m47\fss294(10) & $-$47\degs31\amin24\farcs78(15)  & 20  & 153 & 0.07 & 1.1 & 0.15\\
3 & 17\h32\m20\fss880(16) & $-$47\degs42\amin06\farcs04(24)  & 12  & 17  & 0.82& 0.57 & 0.28\\
4 & 17\h32\m38\fss665(12) & $-$47\degs62\amin12\farcs03(17)  & 18  &54 & 0.18& \nodata&\nodata   \\
5 & 17\h31\m41\fss379(7) & $-$47\degs40\amin16\farcs59(10)  & 29 & 0.3& 0.01 &5.7&0.6  \\
6 & 17\h31\m09\fss344(31) & $-$47\degs48\amin54\farcs89(47)  & 7 &1.6 &0.76 &\nodata&\nodata  \\
7 & 17\h31\m57\fss777(9) & $-$47\degs54\amin49\farcs68(13)  & 24 & 123& 0.07 &4.3&1.2 \\
8 & 17\h33\m10\fss537(10) & $-$47\degs52\amin22\farcs49(14)  & 22 & 498 & 0.28 &10.4& 1.8 \\
\enddata
  \tablecomments{Source coordinates are given as measured on the 2011 image relative to the
  respective calibrator. $S/N$ is the source signal to noise ratio on the 2011 image. Four last columns show the weighted squared residuals ($\chi^2$) for individual sources for two image pairs before and after the registration; subscripts 11--16 and 11--05 denote the image pairs of 2011 and 2016, and of
  2011 and 2005, respectively. Sources 4 and 6 were not detected firmly in the 2005 image and were
  not used.
 }
\end{deluxetable*}
Eight firmly detected  point-like sources located in the pulsar vicinity
were used for the relative astrometry of the ATCA images. Source positions on the 2011 image are listed in Table~\ref{tab:t-atca-ref-sour-samesc}. This epoch is used as the reference. Two sources listed in Table~\ref{tab:t-atca-ref-sour-samesc} were not detected with the sufficient $S/N$ in the 2005 image.
We define the discrepancy in the indiviual source positions on two images by means of $\chi^2$ as
\begin{equation}
\chi^2=\Delta x^T W^{-1} \Delta x,
\end{equation}
where $\Delta x = x_2-x_1$, $W=W_1+W_2$, and $(x_i,W_i),\ i=1,2$ are
source positions and their covariance matrices for two epochs 1 and 2.
Then the initial $\chi^2$ summed over the sources is $\sum \chi^2_{11-05}=31$ for 12 dof (number of coordinates) for 2005--2011 astrometry and  $\sum \chi^2_{11-05}=903$ for 16 dof for 2016--2011 astrometry. It is clear    that the 2016 map differs
substantially from the 2011 map. The same is true for the 2005 map, but to the less extent.
Therefore we matched the 2005 and 2016 images to the 2011 image allowing for the coordinate
transformation in the form
\begin{equation}\label{eq:transform}
\mathbf{x}'=a\hat{R}(\Theta)\left(\mathbf{x}+\delta \mathbf{x} \right),
\end{equation}
where $\mathbf{x}$ is the initial position in the image, $\mathbf{x}'$ is
the transformed position on the reference epoch,
$\delta \mathbf{x}$ is the shift between maps,
$\hat{R}(\Theta)$ is the rotation matrix which rotates the image by the angle $\Theta$, and $a$ is the scale parameter. The source positions on the 2011 epoch were fitted simultaneously with the transformation parameters, and, notably the pulsar position and proper motion. The pulsar positions here, however, were measured on the same images and using the same reference sources, i.e. in the off-pulse (unbinnned) mode in contrast to the positions used for the I solution in Section~\ref{S:atca}.
\begin{widetext}
\begin{deluxetable*}{*{7}{l}}[h]
\tablecaption{Astrometric solution with respect to the 2011 image.\label{tab:t-relastr-samesc}}
\tablehead{
\colhead{Epoch}  &  \colhead{$\delta x_\alpha$} & \colhead{$\delta x_\delta$} & \colhead{$\Theta$}  & \colhead{$a-1$}  &\colhead{$\chi^2_\mathrm{tot}$} &\colhead{$\chi^2_\mathrm{tot}$}
 \\
  \colhead{} & \colhead{asec}& \colhead{asec} & \colhead{amin}  & \colhead{$10^{-3}$}  &\colhead{before}& \colhead{after}
}
\startdata
2005 & $-0.59\pm0.16$ & $0.07\pm 0.15$  &  $-2.0\pm0.5$ & $2.1\pm 0.1$ & 31 & 5.7 \\
2016 & $0.03\pm 0.06$ & $-0.56\pm 0.11$  &  $-0.2\pm0.5$  &  $=a_{2005}$ & 903 & 2.6\\
\enddata
 \tablecomments{$\delta x_\alpha$ and $\delta x_\delta$ are the image shifts in R.A. and decl. directions, respectively, $\Theta$ is the rotation angle, while the scale parameter $a$ is set the same for 2005$-$2011 and 2016$-$2011 registrations. Two last columns show the total $\chi^2$ before and after the registration (sums of the corresponding columns in Table~\ref{tab:t-atca-ref-sour-samesc}).
}
\end{deluxetable*}
\end{widetext}
Initially we performed registration allowing for the linear
transformations only, i.e. setting $a=1$ in
Equation~(\ref{eq:transform}). Although sources' offsets had
generally reduced, there were still large differences in
2011--2016 positions (total discrepancy  $\sum
\chi^2_{11-16}=78.8$), while the 2005--2011 registration was
formally good, with $\sum \chi^2_{11-05}=8.7$. These values and
the visual inspection of the shifted source positions suggested
that the linear transformation was not enough. Allowing the scale
parameters to vary as well, we got a much better solution.
Moreover, we found that the scale parameters for 2005 and 2016
epochs are comparable within uncertainties suggesting that it is
actually the 2011 map that needs to be scaled. Therefore we
performed the final registration tying the scale parameters for
2005--2011 and 2016--2011 registrations. The source registration
becomes fairly good with $\sum \chi^2_{11-16}=2.6$ and $\sum
\chi^2_{11-05}=5.7$ as indicated in
Table~\ref{tab:t-relastr-samesc} where the best-fit astrometric
solution parameters are given.  Examining  these parameters  we
conclude that there is a significant (5$\sigma$) shift between the
2011 and 2016 images in decl. direction and the R.A. shift (at
3.7$\sigma$ level) between 2005 and 2011 images. In addition, the
small rotation is preferrable between the 2005 and 2011 maps. The
possible cause of the rotation can be due to one correlator cycle
(10~s) timing mismatch between $uvw$ and visibiltities. The scale
parameter $a=1.0021(1)$ is well-determined indicating that there
is a problem in a 2011 image reconstruction. A possible cause of
this systematics can be additional frequency averaging performed
in the correlator control computer of CABB in the pulsar binning
mode, a setup specific to the 2011 epoch. The pulsar proper motion
determined in this solution is $\mu_\alpha=81\pm19$ mas~yr$^{-1}$
and $\mu_\delta=-132\pm23$ mas~yr$^{-1}$. We adopt these values as
our `relative astrometry' estimate in
Sections~\ref{S:atca}--\ref{S:comb}. We also checked that the
solution that completely neglects the rotation is slightly worse,
but acceptable and results in the same p.m. parameters within
errors.

\newpage
\bibliographystyle{aasjournal}

\begin{thebibliography}{}
\expandafter\ifx\csname natexlab\endcsname\relax\def\natexlab#1{#1}\fi
\providecommand{\url}[1]{\href{#1}{#1}}
\providecommand{\dodoi}[1]{doi:~\href{http://doi.org/#1}{\nolinkurl{#1}}}
\providecommand{\doeprint}[1]{\href{http://ascl.net/#1}{\nolinkurl{http://ascl.net/#1}}}
\providecommand{\doarXiv}[1]{\href{https://arxiv.org/abs/#1}{\nolinkurl{https://arxiv.org/abs/#1}}}

\bibitem[{{Astropy Collaboration} {et~al.}(2013){Astropy Collaboration},
  {Robitaille}, {Tollerud}, {Greenfield}, {Droettboom}, {Bray}, {Aldcroft},
  {Davis}, {Ginsburg}, {Price-Whelan}, {Kerzendorf}, {Conley}, {Crighton},
  {Barbary}, {Muna}, {Ferguson}, {Grollier}, {Parikh}, {Nair}, {Unther},
  {Deil}, {Woillez}, {Conseil}, {Kramer}, {Turner}, {Singer}, {Fox}, {Weaver},
  {Zabalza}, {Edwards}, {Azalee Bostroem}, {Burke}, {Casey}, {Crawford},
  {Dencheva}, {Ely}, {Jenness}, {Labrie}, {Lim}, {Pierfederici}, {Pontzen},
  {Ptak}, {Refsdal}, {Servillat}, \& {Streicher}}]{Astropy2013A&A}
{Astropy Collaboration}, {Robitaille}, T.~P., {Tollerud}, E.~J., {et~al.} 2013,
  \aap, 558, A33, \dodoi{10.1051/0004-6361/201322068}

\bibitem[{{Astropy Collaboration} {et~al.}(2018){Astropy Collaboration},
  {Price-Whelan}, {Sip{\H o}cz}, {G{\"u}nther}, {Lim}, {Crawford}, {Conseil},
  {Shupe}, {Craig}, {Dencheva}, {Ginsburg}, {VanderPlas}, {Bradley},
  {P{\'e}rez-Su{\'a}rez}, {de Val-Borro}, {Aldcroft}, {Cruz}, {Robitaille},
  {Tollerud}, {Ardelean}, {Babej}, {Bach}, {Bachetti}, {Bakanov}, {Bamford},
  {Barentsen}, {Barmby}, {Baumbach}, {Berry}, {Biscani}, {Boquien}, {Bostroem},
  {Bouma}, {Brammer}, {Bray}, {Breytenbach}, {Buddelmeijer}, {Burke},
  {Calderone}, {Cano Rodr{\'{\i}}guez}, {Cara}, {Cardoso}, {Cheedella},
  {Copin}, {Corrales}, {Crichton}, {D'Avella}, {Deil}, {Depagne}, {Dietrich},
  {Donath}, {Droettboom}, {Earl}, {Erben}, {Fabbro}, {Ferreira}, {Finethy},
  {Fox}, {Garrison}, {Gibbons}, {Goldstein}, {Gommers}, {Greco}, {Greenfield},
  {Groener}, {Grollier}, {Hagen}, {Hirst}, {Homeier}, {Horton}, {Hosseinzadeh},
  {Hu}, {Hunkeler}, {Ivezi{\'c}}, {Jain}, {Jenness}, {Kanarek}, {Kendrew},
  {Kern}, {Kerzendorf}, {Khvalko}, {King}, {Kirkby}, {Kulkarni}, {Kumar},
  {Lee}, {Lenz}, {Littlefair}, {Ma}, {Macleod}, {Mastropietro}, {McCully},
  {Montagnac}, {Morris}, {Mueller}, {Mumford}, {Muna}, {Murphy}, {Nelson},
  {Nguyen}, {Ninan}, {N{\"o}the}, {Ogaz}, {Oh}, {Parejko}, {Parley}, {Pascual},
  {Patil}, {Patil}, {Plunkett}, {Prochaska}, {Rastogi}, {Reddy Janga},
  {Sabater}, {Sakurikar}, {Seifert}, {Sherbert}, {Sherwood-Taylor}, {Shih},
  {Sick}, {Silbiger}, {Singanamalla}, {Singer}, {Sladen}, {Sooley},
  {Sornarajah}, {Streicher}, {Teuben}, {Thomas}, {Tremblay}, {Turner},
  {Terr{\'o}n}, {van Kerkwijk}, {de la Vega}, {Watkins}, {Weaver}, {Whitmore},
  {Woillez}, {Zabalza}, \& {Astropy Contributors}}]{Astropy2018AJ}
{Astropy Collaboration}, {Price-Whelan}, A.~M., {Sip{\H o}cz}, B.~M., {et~al.}
  2018, \aj, 156, 123, \dodoi{10.3847/1538-3881/aabc4f}

\bibitem[{{Bailer-Jones} {et~al.}(2018){Bailer-Jones}, {Rybizki}, {Fouesneau},
  {Mantelet}, \& {Andrae}}]{2018bailer-jones}
{Bailer-Jones}, C.~A.~L., {Rybizki}, J., {Fouesneau}, M., {Mantelet}, G., \&
  {Andrae}, R. 2018, \aj, 156, 58, \dodoi{10.3847/1538-3881/aacb21}

\bibitem[{{Bedford} {et~al.}(1984){Bedford}, {Elliott}, {Ramsey}, \&
  {Meaburn}}]{Bedford1984MNRAS}
{Bedford}, D.~K., {Elliott}, K.~H., {Ramsey}, B., \& {Meaburn}, J. 1984,
  \mnras, 210, 693, \dodoi{10.1093/mnras/210.3.693}

\bibitem[{{Brisken} {et~al.}(2006){Brisken}, {Carrillo-Barrag{\'a}n}, {Kurtz},
  \& {Finley}}]{Brisken2006ApJ}
{Brisken}, W.~F., {Carrillo-Barrag{\'a}n}, M., {Kurtz}, S., \& {Finley}, J.~P.
  2006, \apj, 652, 554, \dodoi{10.1086/507765}

\bibitem[{{Brisken} {et~al.}(2003){Brisken}, {Fruchter}, {Goss}, {Herrnstein},
  \& {Thorsett}}]{brisken2003AJ}
{Brisken}, W.~F., {Fruchter}, A.~S., {Goss}, W.~M., {Herrnstein}, R.~M., \&
  {Thorsett}, S.~E. 2003, \aj, 126, 3090, \dodoi{10.1086/379559}

\bibitem[{{Cappa de Nicolau} {et~al.}(1988){Cappa de Nicolau}, {Niemela},
  {Dubner}, \& {Arnal}}]{Cappa1988AJ}
{Cappa de Nicolau}, C.~E., {Niemela}, V.~S., {Dubner}, G.~M., \& {Arnal}, E.~M.
  1988, \aj, 96, 1671, \dodoi{10.1086/114918}

\bibitem[{{Chatterjee} {et~al.}(2004){Chatterjee}, {Cordes}, {Vlemmings},
  {Arzoumanian}, {Goss}, \& {Lazio}}]{Chatterjee2004ApJ}
{Chatterjee}, S., {Cordes}, J.~M., {Vlemmings}, W.~H.~T., {et~al.} 2004, \apj,
  604, 339, \dodoi{10.1086/381748}

\bibitem[{{Cioffi} {et~al.}(1988){Cioffi}, {McKee}, \&
  {Bertschinger}}]{Cioffi1988ApJ}
{Cioffi}, D.~F., {McKee}, C.~F., \& {Bertschinger}, E. 1988, \apj, 334, 252,
  \dodoi{10.1086/166834}

\bibitem[{{Coles} {et~al.}(2011){Coles}, {Hobbs}, {Champion}, {Manchester}, \&
  {Verbiest}}]{Coles2011MNRAS}
{Coles}, W., {Hobbs}, G., {Champion}, D.~J., {Manchester}, R.~N., \&
  {Verbiest}, J.~P.~W. 2011, \mnras, 418, 561,
  \dodoi{10.1111/j.1365-2966.2011.19505.x}

\bibitem[{{Condon} {et~al.}(1993){Condon}, {Griffith}, \&
  {Wright}}]{1993condon}
{Condon}, J.~J., {Griffith}, M.~R., \& {Wright}, A.~E. 1993, \aj, 106, 1095,
  \dodoi{10.1086/116707}

\bibitem[{{Cordes} \& {Lazio}(2002)}]{cordes2002astro.ph}
{Cordes}, J.~M., \& {Lazio}, T.~J.~W. 2002, ArXiv Astrophysics e-prints

\bibitem[{{D'Alessandro} {et~al.}(1993){D'Alessandro}, {McCulloch}, {King},
  {Hamilton}, \& {McConnell}}]{DAlessandro1993MNRAS}
{D'Alessandro}, F., {McCulloch}, P.~M., {King}, E.~A., {Hamilton}, P.~A., \&
  {McConnell}, D. 1993, \mnras, 261, 883, \dodoi{10.1093/mnras/261.4.883}

\bibitem[{{De Luca}(2017)}]{Deluca2017JPhCS}
{De Luca}, A. 2017, Journal of Physics: Conference Series, 932, 012006,
  \dodoi{10.1088/1742-6596/932/1/012006}

\bibitem[{{Deller} {et~al.}(2018){Deller}, {Goss}, {Brisken}, {Chatterjee},
  {Cordes}, {Janssen}, {Kovalev}, {Lazio}, {Petrov}, {Stappers}, \&
  {Lyne}}]{Deller2018arXiv}
{Deller}, A.~T., {Goss}, W.~M., {Brisken}, W.~F., {et~al.} 2018, arXiv
  e-prints, arXiv:1808.09046.
\newblock \doarXiv{1808.09046}

\bibitem[{{Desvignes} {et~al.}(2016){Desvignes}, {Caballero}, {Lentati},
  {Verbiest}, {Champion}, {Stappers}, {Janssen}, {Lazarus}, {Os{\l}owski},
  {Babak}, {Bassa}, {Brem}, {Burgay}, {Cognard}, {Gair}, {Graikou},
  {Guillemot}, {Hessels}, {Jessner}, {Jordan}, {Karuppusamy}, {Kramer},
  {Lassus}, {Lazaridis}, {Lee}, {Liu}, {Lyne}, {McKee}, {Mingarelli},
  {Perrodin}, {Petiteau}, {Possenti}, {Purver}, {Rosado}, {Sanidas}, {Sesana},
  {Shaifullah}, {Smits}, {Taylor}, {Theureau}, {Tiburzi}, {van Haasteren}, \&
  {Vecchio}}]{Desvignes2016MNRAS}
{Desvignes}, G., {Caballero}, R.~N., {Lentati}, L., {et~al.} 2016, \mnras, 458,
  3341, \dodoi{10.1093/mnras/stw483}

\bibitem[{{Duncan} {et~al.}(1997){Duncan}, {Stewart}, {Haynes}, \&
  {Jones}}]{Duncan1997MNRAS}
{Duncan}, A.~R., {Stewart}, R.~T., {Haynes}, R.~F., \& {Jones}, K.~L. 1997,
  \mnras, 287, 722, \dodoi{10.1093/mnras/287.4.722}

\bibitem[{{Edwards} {et~al.}(2006){Edwards}, {Hobbs}, \&
  {Manchester}}]{Edwards2006MNRASTempo2}
{Edwards}, R.~T., {Hobbs}, G.~B., \& {Manchester}, R.~N. 2006, \mnras, 372,
  1549, \dodoi{10.1111/j.1365-2966.2006.10870.x}

\bibitem[{{Espinoza} {et~al.}(2011){Espinoza}, {Lyne}, {Stappers}, \&
  {Kramer}}]{2011espinoza}
{Espinoza}, C.~M., {Lyne}, A.~G., {Stappers}, B.~W., \& {Kramer}, M. 2011,
  \mnras, 414, 1679, \dodoi{10.1111/j.1365-2966.2011.18503.x}

\bibitem[{{Faucher-Gigu{\`e}re} \& {Kaspi}(2006)}]{Faucher2006ApJ}
{Faucher-Gigu{\`e}re}, C.-A., \& {Kaspi}, V.~M. 2006, \apj, 643, 332,
  \dodoi{10.1086/501516}

\bibitem[{{Feroz} {et~al.}(2009){Feroz}, {Hobson}, \& {Bridges}}]{2009feroz}
{Feroz}, F., {Hobson}, M.~P., \& {Bridges}, M. 2009, \mnras, 398, 1601,
  \dodoi{10.1111/j.1365-2966.2009.14548.x}

\bibitem[{{Fesen} {et~al.}(1985){Fesen}, {Blair}, \& {Kirshner}}]{Fesen1985ApJ}
{Fesen}, R.~A., {Blair}, W.~P., \& {Kirshner}, R.~P. 1985, \apj, 292, 29,
  \dodoi{10.1086/163130}

\bibitem[{{Gaensler} \& {Johnston}(1995)}]{Gaensler1995MNRAS}
{Gaensler}, B.~M., \& {Johnston}, S. 1995, \mnras, 277, 1243,
  \dodoi{10.1093/mnras/277.4.1243}

\bibitem[{{Gaustad} {et~al.}(2001){Gaustad}, {McCullough}, {Rosing}, \& {Van
  Buren}}]{2001gaustad}
{Gaustad}, J.~E., {McCullough}, P.~R., {Rosing}, W., \& {Van Buren}, D. 2001,
  \pasp, 113, 1326, \dodoi{10.1086/323969}

\bibitem[{{Gooch}(1996)}]{gooch}
{Gooch}, R. 1996, in Astronomical Society of the Pacific Conference Series,
  Vol. 101, Astronomical Data Analysis Software and Systems V, ed. G.~H.
  {Jacoby} \& J.~{Barnes}, 80

\bibitem[{{Gotthelf} {et~al.}(2013){Gotthelf}, {Halpern}, \&
  {Alford}}]{Gotthelf2013ApJ}
{Gotthelf}, E.~V., {Halpern}, J.~P., \& {Alford}, J. 2013, \apj, 765, 58,
  \dodoi{10.1088/0004-637X/765/1/58}

\bibitem[{{Green}(2014{\natexlab{a}})}]{Green2014}
{Green}, D.~A. 2014{\natexlab{a}}, Bulletin of the Astronomical Society of
  India, 42, 47.
\newblock \doarXiv{1409.0637}

\bibitem[{{Green}(2014{\natexlab{b}})}]{Greenrep2014}
---. 2014{\natexlab{b}}, A Catalogue of Galactic Supernova Remnants (2014 May
  version), available at ``http://www.mrao.cam.ac.uk/surveys/snrs/''

\bibitem[{{Gvaramadze}(2002)}]{Gvaramadze2002}
{Gvaramadze}, V.~V. 2002, ArXiv Astrophysics e-prints

\bibitem[{{Halpern} \& {Gotthelf}(2015)}]{Halpern2015ApJ}
{Halpern}, J.~P., \& {Gotthelf}, E.~V. 2015, \apj, 812, 61,
  \dodoi{10.1088/0004-637X/812/1/61}

\bibitem[{{Haskell} \& {Melatos}(2015)}]{Haskell2015IJMPD}
{Haskell}, B., \& {Melatos}, A. 2015, International Journal of Modern Physics
  D, 24, 1530008, \dodoi{10.1142/S0218271815300086}

\bibitem[{{Heger} {et~al.}(2003){Heger}, {Fryer}, {Woosley}, {Langer}, \&
  {Hartmann}}]{2003heger}
{Heger}, A., {Fryer}, C.~L., {Woosley}, S.~E., {Langer}, N., \& {Hartmann},
  D.~H. 2003, \apj, 591, 288, \dodoi{10.1086/375341}

\bibitem[{{Hobbs} {et~al.}(2005){Hobbs}, {Lorimer}, {Lyne}, \&
  {Kramer}}]{hobbs2005MNRAS}
{Hobbs}, G., {Lorimer}, D.~R., {Lyne}, A.~G., \& {Kramer}, M. 2005, \mnras,
  360, 974, \dodoi{10.1111/j.1365-2966.2005.09087.x}

\bibitem[{{Hobbs} {et~al.}(2006){Hobbs}, {Edwards}, \&
  {Manchester}}]{Hobbs2006MNRASTempo2}
{Hobbs}, G.~B., {Edwards}, R.~T., \& {Manchester}, R.~N. 2006, \mnras, 369,
  655, \dodoi{10.1111/j.1365-2966.2006.10302.x}

\bibitem[{{Hotan} {et~al.}(2004){Hotan}, {van Straten}, \&
  {Manchester}}]{Hotan2004}
{Hotan}, A.~W., {van Straten}, W., \& {Manchester}, R.~N. 2004, \pasa, 21, 302,
  \dodoi{10.1071/AS04022}

\bibitem[{Hunter(2007)}]{Hunter:2007}
Hunter, J.~D. 2007, Computing In Science \& Engineering, 9, 90,
  \dodoi{10.1109/MCSE.2007.55}

\bibitem[{{Igoshev} \& {Popov}(2013)}]{IgoshevPopov2013MNRAS}
{Igoshev}, A.~P., \& {Popov}, S.~B. 2013, \mnras, 432, 967,
  \dodoi{10.1093/mnras/stt519}

\bibitem[{{Janka}(2012)}]{Janka2012ARNPS}
{Janka}, H.-T. 2012, Annual Review of Nuclear and Particle Science, 62, 407,
  \dodoi{10.1146/annurev-nucl-102711-094901}

\bibitem[{{Janka}(2017)}]{Janka2017ApJ}
---. 2017, \apj, 837, 84, \dodoi{10.3847/1538-4357/aa618e}

\bibitem[{{Jankowski} {et~al.}(2017){Jankowski}, {Bailes}, {Barr}, {Bateman},
  {Bhandari}, {Caleb}, {Deller}, {Campbell-Wilson}, {Farah}, {Flynn}, {Green},
  {Hunstead}, {Jameson}, {Keane}, {Krishnan}, {Plant}, {O'Neill}, {Oslowski},
  {Parthasarathy}, {Ravi}, \& {Temby}}]{Jankowski2017ATel}
{Jankowski}, F., {Bailes}, M., {Barr}, E.~D., {et~al.} 2017, The Astronomer's
  Telegram, 10770

\bibitem[{{Jankowski} {et~al.}(2019){Jankowski}, {Bailes}, {van Straten},
  {Keane}, {Flynn}, {Barr}, {Bateman}, {Bhandari}, {Caleb}, {Campbell-Wilson},
  {Farah}, {Green}, {Hunstead}, {Jameson}, {Os{\l}owski}, {Parthasarathy},
  {Rosado}, \& {Venkatraman Krishnan}}]{Jankowski2019MNRAS}
{Jankowski}, F., {Bailes}, M., {van Straten}, W., {et~al.} 2019, \mnras, 484,
  3691, \dodoi{10.1093/mnras/sty3390}

\bibitem[{{Johnston} {et~al.}(1998){Johnston}, {Nicastro}, \&
  {Koribalski}}]{Johnston1998MNRAS}
{Johnston}, S., {Nicastro}, L., \& {Koribalski}, B. 1998, \mnras, 297, 108,
  \dodoi{10.1046/j.1365-8711.1998.01461.x}

\bibitem[{Jones {et~al.}(2001--)Jones, Oliphant, Peterson, {et~al.}}]{Scipy}
Jones, E., Oliphant, T., Peterson, P., {et~al.} 2001--, {SciPy}: Open source
  scientific tools for {Python}.
\newblock \url{http://www.scipy.org/}

\bibitem[{{Kalberla} \& {Haud}(2015)}]{Kalberla2015AsAp}
{Kalberla}, P.~M.~W., \& {Haud}, U. 2015, \aap, 578, A78,
  \dodoi{10.1051/0004-6361/201525859}

\bibitem[{{Kaspi}(1996)}]{Kaspi1996ASPC}
{Kaspi}, V.~M. 1996, in Astronomical Society of the Pacific Conference Series,
  Vol. 105, IAU Colloq. 160: Pulsars: Problems and Progress, ed. S.~{Johnston},
  M.~A. {Walker}, \& M.~{Bailes}, 375

\bibitem[{{Kaspi} {et~al.}(1994){Kaspi}, {Manchester}, {Siegman}, {Johnston},
  \& {Lyne}}]{Kaspi1994ApJ}
{Kaspi}, V.~M., {Manchester}, R.~N., {Siegman}, B., {Johnston}, S., \& {Lyne},
  A.~G. 1994, \apjl, 422, L83, \dodoi{10.1086/187218}

\bibitem[{{Kim} {et~al.}(2010){Kim}, {Min}, {Seon}, {Han}, \&
  {Edelstein}}]{Kim2010ApJ}
{Kim}, I.-J., {Min}, K.-W., {Seon}, K.-I., {Han}, W., \& {Edelstein}, J. 2010,
  \apj, 709, 823, \dodoi{10.1088/0004-637X/709/2/823}

\bibitem[{{Kirichenko} {et~al.}(2015){Kirichenko}, {Shibanov}, {Shternin},
  {Johnston}, {Voronkov}, {Danilenko}, {Barsukov}, {Lai}, \&
  {Zyuzin}}]{Kirichenko2015MNRAS}
{Kirichenko}, A., {Shibanov}, Y., {Shternin}, P., {et~al.} 2015, \mnras, 452,
  3273, \dodoi{10.1093/mnras/stv1420}

\bibitem[{{Kirsten} {et~al.}(2015){Kirsten}, {Vlemmings}, {Campbell}, {Kramer},
  \& {Chatterjee}}]{Kirsten2015AA}
{Kirsten}, F., {Vlemmings}, W., {Campbell}, R.~M., {Kramer}, M., \&
  {Chatterjee}, S. 2015, \aap, 577, A111, \dodoi{10.1051/0004-6361/201425562}

\bibitem[{{Large} {et~al.}(1968){Large}, {Vaughan}, \&
  {Wielebinski}}]{Large1968Natur}
{Large}, M.~I., {Vaughan}, A.~E., \& {Wielebinski}, R. 1968, \nat, 220, 753,
  \dodoi{10.1038/220753a0}

\bibitem[{{Lentati} {et~al.}(2014){Lentati}, {Alexander}, {Hobson}, {Feroz},
  {van Haasteren}, {Lee}, \& {Shannon}}]{Lentati2014MNRAS}
{Lentati}, L., {Alexander}, P., {Hobson}, M.~P., {et~al.} 2014, \mnras, 437,
  3004, \dodoi{10.1093/mnras/stt2122}

\bibitem[{{Li} {et~al.}(2016){Li}, {Wang}, {Yuan}, {Wang}, {Hobbs}, {Lentati},
  \& {Manchester}}]{Li2016MNRAS}
{Li}, L., {Wang}, N., {Yuan}, J.~P., {et~al.} 2016, \mnras, 460, 4011,
  \dodoi{10.1093/mnras/stw1262}

\bibitem[{{Lorimer} \& {Kramer}(2012)}]{Lorimer2012hpa}
{Lorimer}, D.~R., \& {Kramer}, M. 2012, {Handbook of Pulsar Astronomy}
  (Cambridge, UK: Cambridge University Press)

\bibitem[{{Manchester} {et~al.}(2005){Manchester}, {Hobbs}, {Teoh}, \&
  {Hobbs}}]{manchester2005}
{Manchester}, R.~N., {Hobbs}, G.~B., {Teoh}, A., \& {Hobbs}, M. 2005, \aj, 129,
  1993, \dodoi{10.1086/428488}

\bibitem[{{Manchester} {et~al.}(1983){Manchester}, {Newton}, {Goss}, \&
  {Hamilton}}]{Manchester1983MNRAS}
{Manchester}, R.~N., {Newton}, L.~M., {Goss}, W.~M., \& {Hamilton}, P.~A. 1983,
  \mnras, 202, 269, \dodoi{10.1093/mnras/202.2.269}

\bibitem[{{Matthews} {et~al.}(1998){Matthews}, {Wallace}, \&
  {Taylor}}]{Matthews1998ApJ}
{Matthews}, B.~C., {Wallace}, B.~J., \& {Taylor}, A.~R. 1998, \apj, 493, 312,
  \dodoi{10.1086/305112}

\bibitem[{{Meaburn} {et~al.}(1991){Meaburn}, {Goudis}, {Solomos}, \&
  {Laspias}}]{Meaburn1991}
{Meaburn}, J., {Goudis}, C., {Solomos}, N., \& {Laspias}, V. 1991, \aap, 252,
  291

\bibitem[{{Melatos} {et~al.}(2008){Melatos}, {Peralta}, \&
  {Wyithe}}]{Melatos2008ApJ}
{Melatos}, A., {Peralta}, C., \& {Wyithe}, J.~S.~B. 2008, \apj, 672, 1103,
  \dodoi{10.1086/523349}

\bibitem[{{M{\"u}ller}(2017)}]{Muller2017IAUS}
{M{\"u}ller}, B. 2017, in IAU Symposium, Vol. 329, The Lives and Death-Throes
  of Massive Stars, ed. J.~J. {Eldridge}, J.~C. {Bray}, L.~A.~S. {McClelland},
  \& L.~{Xiao}, 17--24

\bibitem[{{M{\"u}ller} {et~al.}(2016){M{\"u}ller}, {Heger}, {Liptai}, \&
  {Cameron}}]{Muller2016MNRAS}
{M{\"u}ller}, B., {Heger}, A., {Liptai}, D., \& {Cameron}, J.~B. 2016, \mnras,
  460, 742, \dodoi{10.1093/mnras/stw1083}

\bibitem[{{M{\"u}ller} {et~al.}(2019){M{\"u}ller}, {Tauris}, {Heger},
  {Banerjee}, {Qian}, {Powell}, {Chan}, {Gay}, \& {Langer}}]{Muller2019MNRAS}
{M{\"u}ller}, B., {Tauris}, T.~M., {Heger}, A., {et~al.} 2019, \mnras, 484,
  3307, \dodoi{10.1093/mnras/stz216}

\bibitem[{{Noutsos} {et~al.}(2013){Noutsos}, {Schnitzeler}, {Keane}, {Kramer},
  \& {Johnston}}]{Noutsos2013MNRAS}
{Noutsos}, A., {Schnitzeler}, D.~H.~F.~M., {Keane}, E.~F., {Kramer}, M., \&
  {Johnston}, S. 2013, \mnras, 430, 2281, \dodoi{10.1093/mnras/stt047}

\bibitem[{{Pejcha} \& {Thompson}(2015)}]{Pejcha2015ApJ}
{Pejcha}, O., \& {Thompson}, T.~A. 2015, \apj, 801, 90,
  \dodoi{10.1088/0004-637X/801/2/90}

\bibitem[{{Popov} \& {Turolla}(2012)}]{PopovTurolla2012Ap&SS}
{Popov}, S.~B., \& {Turolla}, R. 2012, \apss, 341, 457,
  \dodoi{10.1007/s10509-012-1100-z}

\bibitem[{{Ren} {et~al.}(2018){Ren}, {Liu}, {Chen}, {Xiang}, {Yuan}, {Huang},
  {Zhang}, {Wang}, {Tian}, {Liu}, \& {Wu}}]{Ren2018}
{Ren}, J.-J., {Liu}, X.-W., {Chen}, B.-Q., {et~al.} 2018, Research in Astronomy
  and Astrophysics, 18, 111, \dodoi{10.1088/1674-4527/18/9/111}

\bibitem[{{Sault} {et~al.}(1995){Sault}, {Teuben}, \& {Wright}}]{miriad}
{Sault}, R.~J., {Teuben}, P.~J., \& {Wright}, M.~C.~H. 1995, in Astronomical
  Society of the Pacific Conference Series, Vol.~77, Astronomical Data Analysis
  Software and Systems IV, ed. R.~A. {Shaw}, H.~E. {Payne}, \& J.~J.~E.
  {Hayes}, 433

\bibitem[{{Shelton}(1999)}]{1999Shelton}
{Shelton}, R.~L. 1999, \apj, 521, 217, \dodoi{10.1086/307553}

\bibitem[{{Shternin} {et~al.}(2017){Shternin}, {Yu}, {Kirichenko}, {Shibanov},
  {Danilenko}, {Voronkov}, \& {Zyuzin}}]{2017peter}
{Shternin}, P.~S., {Yu}, M., {Kirichenko}, A.~Y., {et~al.} 2017, Journal of
  Physics : Conference Series, 932, 012004,
  \dodoi{10.1088/1742-6596/932/1/012004}

\bibitem[{{Smith} {et~al.}(2019){Smith}, {Bruel}, {Cognard}, {Cameron},
  {Camilo}, {Dai}, {Guillemot}, {Johnson}, {Johnston}, {Keith}, {Kerr},
  {Kramer}, {Lyne}, {Manchester}, {Shannon}, {Sobey}, {Stappers}, \&
  {Weltevrede}}]{Smith2019}
{Smith}, D.~A., {Bruel}, P., {Cognard}, I., {et~al.} 2019, \apj, 871, 78,
  \dodoi{10.3847/1538-4357/aaf57d}

\bibitem[{{Sukhbold} {et~al.}(2016){Sukhbold}, {Ertl}, {Woosley}, {Brown}, \&
  {Janka}}]{Sukhbold2016ApJ}
{Sukhbold}, T., {Ertl}, T., {Woosley}, S.~E., {Brown}, J.~M., \& {Janka}, H.-T.
  2016, \apj, 821, 38, \dodoi{10.3847/0004-637X/821/1/38}

\bibitem[{{Thorsett} {et~al.}(2003){Thorsett}, {Benjamin}, {Brisken}, {Golden},
  \& {Goss}}]{Thorsett2003ApJ}
{Thorsett}, S.~E., {Benjamin}, R.~A., {Brisken}, W.~F., {Golden}, A., \&
  {Goss}, W.~M. 2003, \apjl, 592, L71, \dodoi{10.1086/377682}

\bibitem[{{Vaughan} {et~al.}(1974){Vaughan}, {Nicholson}, \&
  {Disney}}]{Vaughan1974MNRAS}
{Vaughan}, A.~E., {Nicholson}, M.~J., \& {Disney}, P. 1974, \mnras, 168, 361,
  \dodoi{10.1093/mnras/168.2.361}

\bibitem[{{Verbunt} {et~al.}(2017){Verbunt}, {Igoshev}, \&
  {Cator}}]{Verbunt2017A&A}
{Verbunt}, F., {Igoshev}, A., \& {Cator}, E. 2017, \aap, 608, A57,
  \dodoi{10.1051/0004-6361/201731518}

\bibitem[{{Walker} \& {Zealey}(2001)}]{Walker2001MNRAS}
{Walker}, A.~J., \& {Zealey}, W.~J. 2001, \mnras, 325, 287,
  \dodoi{10.1046/j.1365-8711.2001.04423.x}

\bibitem[{{Wang} {et~al.}(2000){Wang}, {Manchester}, {Pace}, {Bailes}, {Kaspi},
  {Stappers}, \& {Lyne}}]{Wang2000MNRAS}
{Wang}, N., {Manchester}, R.~N., {Pace}, R.~T., {et~al.} 2000, \mnras, 317,
  843, \dodoi{10.1046/j.1365-8711.2000.03713.x}

\bibitem[{{Welsh} {et~al.}(2003){Welsh}, {Sallmen}, {Jelinsky}, \&
  {Lallement}}]{Welsh2003}
{Welsh}, B.~Y., {Sallmen}, S., {Jelinsky}, S., \& {Lallement}, R. 2003, \aap,
  403, 605, \dodoi{10.1051/0004-6361:20030168}

\bibitem[{{Wilson} {et~al.}(2011){Wilson}, {Ferris}, {Axtens}, {Brown},
  {Davis}, {Hampson}, {Leach}, {Roberts}, {Saunders}, {Koribalski}, {Caswell},
  {Lenc}, {Stevens}, {Voronkov}, {Wieringa}, {Brooks}, {Edwards}, {Ekers},
  {Emonts}, {Hindson}, {Johnston}, {Maddison}, {Mahony}, {Malu}, {Massardi},
  {Mao}, {McConnell}, {Norris}, {Schnitzeler}, {Subrahmanyan}, {Urquhart},
  {Thompson}, \& {Wark}}]{wilson}
{Wilson}, W.~E., {Ferris}, R.~H., {Axtens}, P., {et~al.} 2011, \mnras, 416,
  832, \dodoi{10.1111/j.1365-2966.2011.19054.x}

\bibitem[{{Wongwathanarat} {et~al.}(2013){Wongwathanarat}, {Janka}, \&
  {M{\"u}ller}}]{Wongwathanarat2013A&A}
{Wongwathanarat}, A., {Janka}, H.~T., \& {M{\"u}ller}, E. 2013, \aap, 552,
  A126, \dodoi{10.1051/0004-6361/201220636}

\bibitem[{{Yao} {et~al.}(2017){Yao}, {Manchester}, \& {Wang}}]{yao2017}
{Yao}, J.~M., {Manchester}, R.~N., \& {Wang}, N. 2017, \apj, 835, 29,
  \dodoi{10.3847/1538-4357/835/1/29}

\bibitem[{{Yar-Uyaniker} {et~al.}(2004){Yar-Uyaniker}, {Uyaniker}, \&
  {Kothes}}]{YarUyaniker2004ApJ}
{Yar-Uyaniker}, A., {Uyaniker}, B., \& {Kothes}, R. 2004, \apj, 616, 247,
  \dodoi{10.1086/424794}

\bibitem[{{Yu} {et~al.}(2013){Yu}, {Manchester}, {Hobbs}, {Johnston}, {Kaspi},
  {Keith}, {Lyne}, {Qiao}, {Ravi}, {Sarkissian}, {Shannon}, \&
  {Xu}}]{Yu2013MNRAS}
{Yu}, M., {Manchester}, R.~N., {Hobbs}, G., {et~al.} 2013, \mnras, 429, 688,
  \dodoi{10.1093/mnras/sts366}

\end{thebibliography}

\clearpage

\end{document}